\documentclass[lettersize,journal]{IEEEtran}
\usepackage{amsmath,amsfonts}
\usepackage[linesnumbered,ruled,vlined]{algorithm2e}
\usepackage{array}
\usepackage{cuted}
 \usepackage{booktabs} 
\usepackage{multirow}
\usepackage{textcomp}
\usepackage{stfloats}
\usepackage{url}
\usepackage{verbatim}
\usepackage{graphicx}
\usepackage{cite}
\usepackage{subcaption} 
\usepackage{newtxtext,newtxmath} 
\usepackage{float} 
\usepackage[hidelinks]{hyperref}

\usepackage{caption}
\begin{document}

\title{Confidence-Guided Cross-Modal Knowledge Transfer for Multimodal Anomaly Detection in Microservice Systems}

\author{Peipeng Wang, Xiuguo Zhang\IEEEauthorrefmark{1}, Lihua Zhang,
Zhiying Cao\IEEEauthorrefmark{1},
Jingchun Zhou, \IEEEmembership{Senior Member, IEEE},
Zheng Li, \IEEEmembership{Senior Member, IEEE}\\

\thanks{Peipeng Wang, Xiuguo Zhang, Lihua Zhang, Zhiying Cao, and Jingchun Zhou are with Dalian Maritime University, Liaoning, Dalian, 116026, China. E-mail: {\{wpp7, zhangxg, zhang\_lihua, czysophy\}@dlmu.edu.cn, zhoujingchun03@gmail.com.}}
\thanks{Zheng Li is with Queen's University Belfast, BT9 5AF Belfast, UK. E-mail: zheng.li@qub.ac.uk.}
\thanks{* Xiuguo Zhang and Zhiying Cao are the corresponding authors.}
}

\markboth{Journal of \LaTeX\ Class Files,~Vol.~14, No.~8, August~2021}%
{Shell \MakeLowercase{\textit{et al.}}: A Sample Article Using IEEEtran.cls for IEEE Journals}


\maketitle

\begin{abstract}
Accurate anomaly detection is essential for reliable and secure operations of microservice systems. While an increasing number of studies have shifted from unimodal modeling to multimodal interaction and fusion, effectively leveraging reliable cross-modal information remains challenging. The challenge primarily stems from two aspects. Firstly, different modalities are influenced by factors like load fluctuations, leading to dynamically changing reliability. Secondly, multimodal data exhibit heterogeneity in both structure and semantics. Therefore, we propose a confidence-guided Cross-Modal knowledge Transfer method for multimodal Anomaly Detection (CMT-AD). It jointly models metrics and logs within a unified deep clustering framework and estimates modality reliability through the soft clustering distributions, where clustering uncertainty is quantified into confidence scores. Guided by these confidence scores, the model actively analyzes the contributions of each modality in cross-modal interactions and supplements low-confidence modalities with knowledge from high-confidence ones. To further mitigate cross-modal heterogeneity, we introduce a gated intermediate modality and design structural and semantic consistency constraints that align the original modalities with the intermediate modality to preserve similarity structures and semantic distributions across modalities. Furthermore, intra-modal and cross-modal regularization terms are incorporated to enhance cluster compactness and mitigate negative transfer. We evaluated CMT-AD on three large-scale datasets, and the results demonstrate that CMT-AD outperforms state-of-the-art approaches and achieves an F1-score higher than 0.9. 
\end{abstract}
\begingroup
\renewcommand\thefootnote{}
\footnotetext{This work has been submitted to the IEEE for possible publication.
Copyright may be transferred without notice, after which this version may no longer be accessible.}
\addtocounter{footnote}{-1}
\endgroup

\section{Introduction}
\IEEEPARstart{I}{n} recent years, microservice architecture has been widely adopted in enterprise applications. As system scale continues to expand, functional structures have become increasingly complex, and dependencies between different microservices have grown tighter~\cite{Moens}. While the service decoupling and flexible composition architectural pattern enhances system scalability and development efficiency, it also makes overall stability more dependent on the health of each individual service. Therefore, to ensure business continuity, service providers must prioritize improving system reliability and stability. Building efficient and automated anomaly detection methods is fundamental to achieving this goal~\cite{DALAD}. 

In microservice systems, the operational status is typically recorded through various types of monitoring data. Among these, performance metrics and log data, representing system states and discrete events, respectively, are the two most fundamental forms of representation. Some studies analyze the temporal features of metrics while incorporating structural dependencies between different metrics~\cite{Yahya,Khettaf}. Others treat logs as natural language, employing word embedding models to extract log semantics~\cite{loggt} and temporal models to capture contextual information~\cite{swisslog}. However, unimodal approaches struggle to comprehensively detect all system anomalies due to the inherent limitations of relying on a single type of monitoring data. As shown in Table~\ref{tab1}, metrics and logs exhibit distinct detection capabilities across different anomaly categories, indicating that their complementary characteristics are essential for achieving more comprehensive and accurate anomaly detection. Recent studies have begun to jointly model metric and log data, enhancing anomaly detection performance through cross-modal interaction and fusion mechanisms~\cite{hades,scwarn}. However, these approaches still struggle to effectively leverage reliable cross-modal information to identify complex anomalies that are difficult to detect using single-modal methods. The main reasons for this include two key points:

\textbf{(1) Reliability of dynamic changes in multimodal data.} In microservice systems, the observability, sensitivity, and discrimination capabilities of different modalities for detecting anomalies are not always consistent. This inconsistency is influenced by the operating mechanisms of the microservice system and the observation mechanisms associated with different modal data. Some anomalies develop through a gradual accumulation and diffusion process, exhibiting distinct behavioral characteristics at various stages. For example, as shown in Table~\ref{tab1}, logs record early warnings such as ``request timeout”, yet these warnings do not necessarily indicate critical failures. As the anomaly evolves, log events may diminish or disappear, while substantial deviations would emerge in performance metrics. Additionally, due to noise and data collection mechanisms, metric values may fluctuate temporarily, and logs may lack critical information because of rate-limiting issues. This can result in a situation where one modality struggles to provide exception clues at a given moment, while another modality proves more reliable. Existing methods overlook this dynamic reliability, leading to information conflicts and dilution of valuable data after fusion or interaction, which causes cross-modal information distortion.

\begin{table}[!t]
\caption{ANOMALY CHARACTERISTICS IN DIFFERENT MODALITIES}\label{tab1}
\centering
\begin{tabular}{|c|c|c|}
\hline
\renewcommand{\arraystretch}{1.3}
Anomaly types & Metric & Log\\ \hline
Thread blocking & Latency↑& -- \\
Request timed out& Timeout Rate↑ & Warning request timed out\\
Login failed& -- & Login failed error\\
Database connection failed & Error rate↑ & Connection failed error\\

\hline
\end{tabular}
\end{table}

\textbf{(2) Heterogeneity in multimodal data structure and semantics.} Metric data consists of multivariate time series with temporal dependencies, continuously collected from the system's operational state. In contrast, log data is generated in an event-driven manner, predefined by the programmer, and populated with specific contextual information at runtime. Logs are semi-structured text representing discrete event sequences~\cite{zhang}. This makes it challenging to directly correlate metrics and logs at the same time step. Some studies have converted logs into time series~\cite{Anofusion,twin} and performed multimodal fusion, but this approach results in the loss of semantic information inherent in the logs. Furthermore, metrics represent the system state through numerical changes, with their semantics being implicit, whereas logs explicitly describe the system state. In anomaly detection scenarios, variations in a specific metric variable do not necessarily indicate a unique anomaly type. For example, a continuous increase in memory usage may be caused by either memory leaks or frequent garbage collection, while logs provide a fine-grained description of explicit events related to these anomalies. This heterogeneity in data structure and semantic expression complicates the direct alignment of information from different modalities within a unified representation space, thereby increasing the difficulty of cross-modal joint analysis.

To address these challenges, we propose CMT-AD, innovatively integrating deep clustering paradigm and cross-modal transfer methods into anomaly detection based on metrics and logs. We first dynamically evaluate sample reliability over time across modalities using the soft clustering assignments, and quantify it as confidence scores. This enables us to supplement low-confidence modalities with knowledge from high-confidence modalities during cross-modal knowledge transfer, adaptively adjusting the amount of transferred information based on confidence differences. Furthermore, an intra-modal regularization term is introduced to bring representations within the same cluster closer together while pushing apart those from different clusters, thereby stabilizing transferable knowledge.

To further mitigate instability introduced by direct cross-modal interaction, we employ gating mechanisms and confidence scores to integrate reliable knowledge from different modalities, creating an intermediate modality. Within this shared representation space, we impose two complementary transfer constraints. The structural consistency constraint preserves the geometric similarity relationships among samples across modalities, while the semantic consistency constraint aligns modality-specific representations with shared semantic prototypes in the intermediate modality. A cross-modal regularization term is designed to preserve discriminative boundaries, thus preventing negative transfer.

The main contributions are summarized as follows:
\begin{itemize}
    \item We introduce deep clustering and cross-modal transfer learning into anomaly detection for microservice systems, enabling unified analysis of metric and log data within an unsupervised framework to enhance detection accuracy and robustness.
    \item We design a confidence estimation method based on soft clustering assignment,to quantify modality reliability over time, mitigating the influence of unreliable modalities. Additionally, an intra-modal regularization term is introduced to promote compact and separable representations within each modality.
    \item We construct an intermediate modality and introduce structural and semantic consistency constraints to alleviate heterogeneity across representations. A cross-modal regularization term is further developed to guide modal interaction and enhance the effectiveness of knowledge transfer.
\end{itemize}

\section{Related Work}

\subsection{Anomaly Detection with Single-modal Data}

\textbf{Metrics:} Existing research treats metrics as multivariate time series and employs deep learning methods to enhance anomaly detection performance. OmniAnomaly~\cite{OmniAnomaly} utilized a GRU-based VAE model to reconstruct the normal state of multivariate time series. AnomalyTrans~\cite{anomalytrans} leveraged Transformer to model time series and introduces an association difference loss function to quantify the disparity between normal and abnormal sequences. Dual-TF~\cite{Nam} incorporated contrastive learning into the Transformer to address the issue of time-frequency granularity mismatch. Several studies have identified correlations between different metric variables. CGAD~\cite{Febrinanto} employed transfer entropy to construct a causal graph among variables and utilized weighted GCN alongside to analyze its structural and temporal relationships. MUTANT~\cite{mutant} also applied GCN and LSTM to learn the correlations between variable. Luo et al.~\cite{Dual} modeled time series features as intra-signal graphs and different variables as inter-signal graphs. Moon~\cite{Moon} introduced a multivariate Markov transfer field technique to transform time series data into graph structures and used multimodal CNNs to analyze their spatiotemporal features. 

\textbf{Logs:} Existing logs anomaly detection approaches typically divide the process into three stages: log parsing, feature extraction, and anomaly detection. Commonly used log parsing methods include Drain~\cite{drain}, IPLoM~\cite{makanju}, and SwissLog~\cite{swisslog}. Zhu et al.~\cite{zhu} demonstrated that Drain can effectively parse complete log events from noisy data. Feature extraction methods are generally categorized into two types: index-based and semantic-based approaches. DeepLog~\cite{deeplog} mapped log events to unique IDs and employed an LSTM model to learn the log index sequence. LogAnomaly~\cite{loganomaly} calculated the frequency of each log index at a given time to obtain statistical features. LogRobust~\cite{logbust} pointed out that the log index ignores semantics and reduces the robustness of anomaly detection models. LogMTC~\cite{he} and DeepSysLog~\cite{deepsyslog} used Word2vect to extract log word vectors, LightLog~\cite{lightlog} combines Word2vec and post-processing algorithms to generate low-dimensional word vectors, Chai~\cite{chai} et al. utilized BERT to obtain log sentence features, and LogGraph~\cite{loggraph} used a pre-trained GloVe model to generate log semantic. 

In the process of anomaly detection, IST-GCN~\cite{XU} employed both directed and undirected graphs to capture the temporal and spatial features of log events. SpikeLog~\cite{SpikeLog} designed a Spiking Neural Network to perform anomaly detection. SLNALog~\cite{SLNALog} utilized linear attention to address the key-value interaction challenges inherent in traditional attention mechanisms. LogDLR~\cite{LogDLR} incorporated adversarial training within a Transformer to learn invariant latent representations from log entries. R-Log~\cite{R-Log} integrated a reinforcement learning inference mechanism into a large language model to improve the adaptability of log analysis methods across different domains. 

These methods analyze metrics and log characteristics from various perspectives, effectively enhancing detection results. However, they remain limited to a single type of monitoring data, which makes it challenging to meet real-world operational requirements.

\subsection{Anomaly Detection with Multimodal Data}

In recent years, an increasing number of studies have combined metrics and logs to enhance anomaly detection performance. SCWarn~\cite{scwarn} transformed logs into time series and employed different LSTM models to learn features from both metrics and log temporal feature. Although it effectively performed joint analysis of metrics and logs, it overlooked cross-modal interactions. Some studies have further incorporated trace data. HG-PAD~\cite{HG-PAD} construct a multirelational heterogeneous graph using three types of monitoring data. FAMOS~\cite{Famos} introduced a novel Gaussian attention mechanism to emphasize the relationships among the three modalities. While these methods leverage traces to enrich multimodal features, trace collection requires pre-instrumentation and incurs higher computational costs compared to logs and metrics. Therefore, this paper focuses on anomaly detection based on metrics and logs. 

KANAD~\cite{kanad} designed a graph structure based on K-nearest neighbors for metrics and a word-level graph structure for logs, achieving feature fusion through cross-modal interaction. Hades~\cite{hades} applied an attention mechanism to facilitate cross-modal interaction. However, these methods typically rely on labeled data for training, limiting their applicability in unsupervised scenarios. To address this, MAD-CMC~\cite{mad-cmc} employed clustering methods and contrastive learning to enhance the model's ability to distinguish samples from different clusters. UAC-AD~\cite{uac-ad} developed an adversarial training mechanism for hard samples in anomaly detection, which automatically identifies these hard samples and adjusts their weights accordingly.

While these methods effectively facilitate cross-modal interactions, they overlook the fact that different modalities exhibit varying reliabilities over time. Furthermore, interaction mechanisms such as attention-based methods, which rely on weighted feature fusion remain susceptible to cross-modal heterogeneity. Therefore, we analyze modality reliability across time and model cross-modal features from both structural and semantic perspectives, thereby mitigating performance degradation caused by heterogeneity.

\section{Motivation}
\textbf{(1) Existing methods rely on the static fusion assumption, which treats knowledge from different modalities as equally reliable during the anomaly detection process.}

Although recent research has shifted from unimodal to multimodal approaches, most interaction mechanisms adopt fixed fusion strategies that ignore the temporal variability of modality reliability. In real-world microservice systems, anomalies evolve progressively, exhibiting stage-dependent characteristics. As a result, different modalities demonstrate varying discrimination capabilities at different stages of anomaly progression. Moreover, data collection noise, sampling inconsistencies, and environmental disturbances may cause certain modalities to temporarily provide incomplete or misleading signals. These factors challenges the effectiveness of static fusion methods. Therefore, we explicitly model modality reliability as a time-dependent confidence score to dynamically adjust cross-modal interactions.

\textbf{(2) Direct cross-modal fusion or alignment overlooks the structural and semantic differences between modalities, leading to unstable interaction and degraded anomaly detection performance.} 

Metric and log data exhibit fundamentally different representational characteristics. Metric data consist of continuous time-series signals that capture system dynamics through smooth temporal variations, whereas log data consist of discrete event entries enriched with high-level semantic information, where keywords such as ``success,” ``request,” and ``error” explicitly reflect underlying system states. Beyond differences in data format, these modalities exist in heterogeneous representational spaces with distinct structural dependencies and semantic abstractions. Existing research often aligns or fuses multimodal features directly using attention mechanisms~\cite{hades,scwarn}, assuming that weighted feature aggregation is sufficient to bridge cross-modal gaps. However, these strategies primarily operate at the feature level and fail to address structural inconsistencies and semantic misalignments across modalities. Such mismatches can distort cross-modal interactions and impede stable knowledge integration. Therefore, we construct an intermediate modality and explicitly impose structural and semantic consistency constraints to standardize representation alignment and achieve stable cross-modal fusion.

\textbf{(3) In complex microservice scenarios, indiscriminate cross-modal knowledge transfer can easily result in negative transfer.}

Existing research on anomaly detection using transfer learning focuses on transferring knowledge or models from a source modality to a target modality to achieve rapid deployment and improved performance~\cite{loggt}. The heterogeneity and reliability differences among modalities are often overlooked. Although the knowledge transfer process resembles cross-modal interaction, not all modal information is suitable for transfer at all times. In microservice systems, when unreliable or poorly aligned modal knowledge is indiscriminately propagated, it can distort feature representations, disrupt modality-specific discrimination capabilities, and ultimately lead to negative transfer. Therefore, effective cross-modal transfer should be selectively regulated rather than indiscriminately enforced. To address this, we propose two regularization terms based on clustering results, targeting both intra-and cross-modal aspects to prevent negative transfer.

\section{Methodology of CMT-AD}
\subsection{Overview}
\begin{figure*}[!t]
    \centering
    \includegraphics[width=0.85\textwidth]{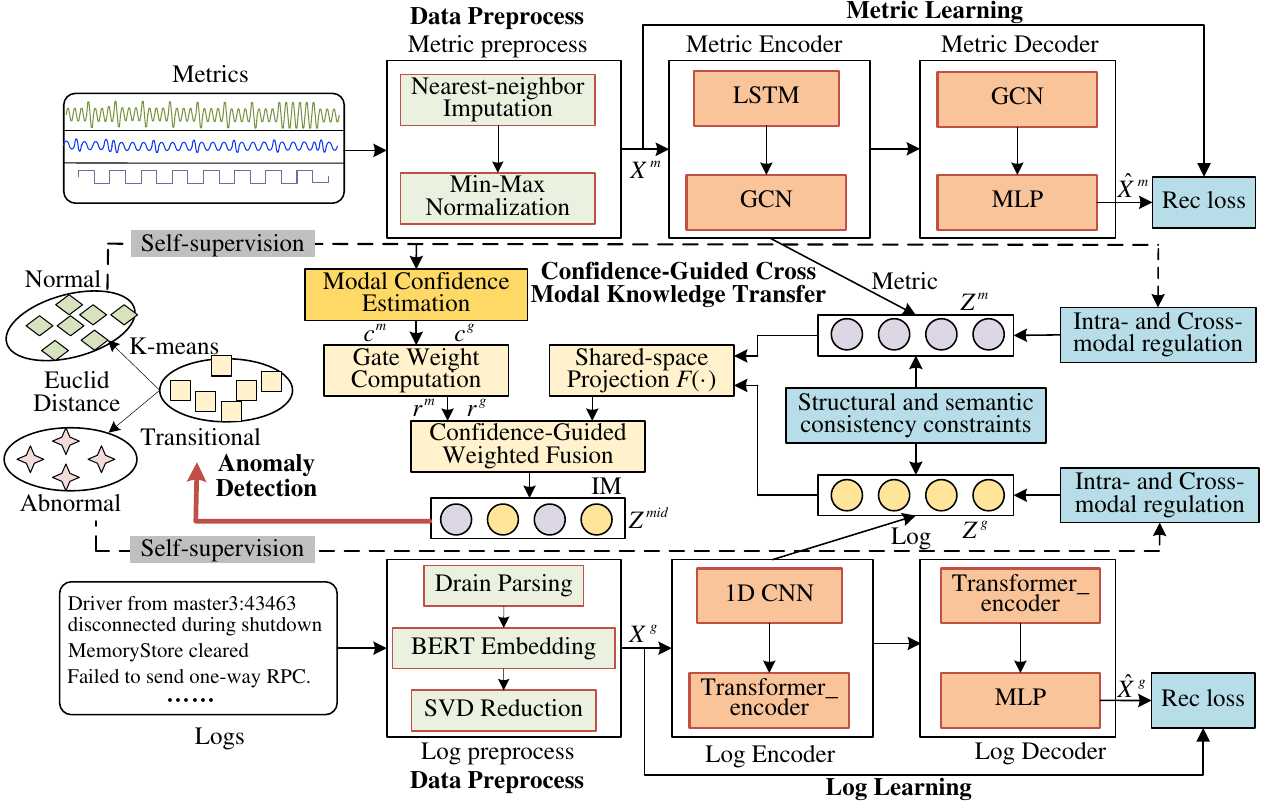}
    \caption{The architecture of CMT-AD.}
    \label{fig1}
\end{figure*}

As shown in the Fig.~\ref{fig1}. CMT-AD primarily consists of five modules: multimodal data preprocessing, metric learning, log learning, confidence-guided cross-modal knowledge transfer, and anomaly detection. In large-scale microservice systems, each microservice can independently generate metrics and logs. Formally, we define the multimodal data as $< {X_{1:n}},{Y_{1:n}} >  = \{ ({X_1},{Y_1}),({X_2},{Y_2}),...,({X_n},{Y_n})\}$, where $n$ represents the data volume. ${X_i}$ contains metrics $X_i^m$ and logs $X_i^g$, and ${Y_i}$ takes values 0 (normal) or 1 (abnormal). To reduce dependence on labeled data, we employ an unsupervised clustering algorithm to classify multimodal monitoring data into different clusters and determine normal and abnormal cluster centers by analyzing these clusters. For new multimodal monitoring data, CMT-AD assigns them to the appropriate cluster to infer the current system status.

\subsection{Multimodal Data Preprocessing}
For metric and log data from microservice systems, we first perform data alignment to ensure that data from different modalities can collaboratively characterize the current system state. Metric data are typically multivariate time series with fixed sampling intervals. Existing research~\cite{OmniAnomaly,Nam} usually segments metrics according to fixed-length time windows. Under the same sampling interval, dozens of log messages may be generated. To align logs with metric windows, these methods~\cite{logbust,loggt} typically aggregate the log messages produced within each fixed time interval into a corresponding log sequence. Therefore, we define a sliding window of size $T$ for the metric data, with the metric sequence at time $i$ as $X_i^m = \{ x_i^m,x_{i - 1}^m,...,x_{i - T - 1}^m\}$, where $x_i^m \in {{\mathbb R}^k}$ and $k$ are the number of variables. Then, we obtain the start and end times of the metric sequence in the current window, and use them to obtain all log messages during the same time period, resulting in the corresponding log sequence $X_i^g = \{ x_{i,1}^g,x_{i,2}^g,...,x_{i,n1}^g\}$, where $n1$ is the number of log messages in the current window.

Subsequently, to enhance the model's robustness and minimize the impact of noisy data, we performed missing value imputation and data normalization on the metric data. For missing value imputation, we applied the nearest neighbor method, filling in missing metric values with valid observations from adjacent time points, thereby preserving the continuity of the metric time series. For data normalization, we used min-max scaling for each dimension of the metric variables to mitigate the effects of differing units on model training.

Since log messages are semi-structured text, we use the Drain algorithm to parse them into structured log events, thereby eliminating redundant information from free text. Then, we input log events into the BERT model~\cite{bert} to obtain semantic features. During the training of CMT-AD, BERT is used as a fixed feature extractor, and its parameters remain frozen during training rather than being fine-tuned end-to-end. However, given the high dimensionality of the semantic features produced by BERT, combining them with metric features during training can cause instability due to scale imbalance. To address this, we apply Singular Value Decomposition (SVD) to reduce the feature dimensionality. The reduced dimension is selected according to the cumulative explained variance criterion, where the retained components preserve at least 95\% of the total variance, thereby retaining the main semantic information while reducing feature redundancy.

\subsection{Metric Learning}
We designed an metric learning module that is pre-trained using training data containing both normal and abnormal samples, based on a reconstruction approach. The goal of this step is to learn a smooth representation space rather than an anomaly discrimination mechanism, thereby providing a stable initialization for subsequent cross-modal analysis and clustering. During encoding, we jointly capture the temporal and spatial dependencies of metric data to comprehensively characterize metric-driven system anomalies. Considering that anomalies in microservice systems evolve gradually over time across each dimension of the metrics, we employ an LSTM to model metric sequences and capture both long- and short-term temporal dependencies:
\begin{equation}
h_t^m = {\mathop{\rm LSTM}\nolimits} (X_t^m)
\label{eq1}
\end{equation}
where $X_t^m \in {{\mathbb R}^{k \times T}}$ represents the metric feature of sliding window at time $t$, $h_t^m \in {{\mathbb R}^{k \times h'}}$ represents the output metric temporal features.

In microservice systems, anomalies often result in correlated changes across multiple metric variables. Therefore, CMT-AD models metric variables using a graph structure, where each variable is represented as a node, and its corresponding temporal features $h_t^m$ are encoded as node embeddings. The cosine similarity is used to construct an adjacency matrix $A \in {{\mathbb R}^{k \times k}}$, the calculation is shown in Eq.~\ref{eq3}.

\begin{equation}
{e_{ij}} = \frac{{h_{t,i}^m \cdot h_{t,j}^m}}{{||h_{t,i}^m|| \times ||h_{t,j}^m||}}
\label{eq3}
\end{equation}
where ${e_{ij}}$ represents the similarity value between the features of nodes $i$ and $j$. When $j \in TopK({e_{ik}})$, ${A_{ij}} = 1$, otherwise ${A_{ij}} = 0$. $TopK({e_{ik}})$ represents the $K$ nodes selected from all nodes that have the highest similarity to node $i$.  This method constructs a metric structure graph and uses the hyperparameter $K$  to control the graph's sparsity, thereby adjusting the graph's complexity across different systems. Subsequently, we employ a two-layer graph convolutional network (GCN) to aggregate local neighborhood information while preserving the graph's structural information, effectively capturing higher-order dependencies among different variables. The calculation is as follows:
\begin{equation}
Z_t^m = \sigma \left( {{{\tilde A}_t}\sigma ({{\tilde A}_t}h_t^m{W^1}){W^2}} \right)\label{eq4}
\end{equation}
where $Z_t^m \in {{\mathbb R}^{k \times {d_m}}}$ contains both temporal and spatial features, ${d_m}$ is the hidden layer dimension of the metric. ${{W^1}}$ and ${{W^2}}$ is learnable weight parameters, ${\tilde A_t}$ is the result of adding a self-loop to the adjacency matrix, ${\tilde A_t} = {A_t} + I$. 

During the decoding stage, the encoded representations $Z_t^m$ are fed into a GCN together with the corresponding adjacency matrix $A_t$, which reinforces the learned structural dependencies and ensures consistency with the graph-based modeling strategy adopted in the encoding phase. Unlike the encoder, the decoder does not perform additional temporal modeling. Instead, it focuses on reconstructing metric representations while preserving structural consistency. Specifically, an MLP layer is applied after the GCN to perform a nonlinear transformation on the structure-enhanced features, producing the reconstructed metric representations. The corresponding formulation is given as follows:
\begin{equation}
\hat X_t^m = {\mathop{\rm MLP}\nolimits} \left( {{\mathop{\rm GCN}\nolimits} (Z_t^m,{A_t})} \right)
\label{eq5}
\end{equation}
where $\hat X_t^m \in {{\mathbb R}^{k \times T}}$ represents reconstructed metric feature. During pre-training, we use the mean squared error loss function to calculate the reconstruction error.

\begin{equation}
er{r^m} = \sum\limits_{t = 1}^n {{{(X_t^m - \hat X_t^m)}^2}}\label{eq7}
\end{equation}

\subsection{Log Learning}
Similar to the metric learning module, the log learning module is pre-trained using a reconstruction approach. Specifically, anomalies may cause recurring adjacent log events within a short time interval, forming abnormal local patterns. To capture such patterns, CMT-AD incorporates a 1D-CNN model during the encoding stage. By applying convolutional operations to the log semantic sequence, it extracts local semantic features from adjacent log events.

\begin{equation}
h_t^g = {\mathop{\rm CNN}\nolimits}(X_t^g)\label{eq8}
\end{equation}
where $X_t^g \in {{\mathbb R}^{s \times n1}}$ denotes the log semantic sequence formed by the semantic representations of log events within the sliding window at time $t$, $s$ represents the log semantic dimension after SVD algorithm. $h_t^g \in {{\mathbb R}^{{d_c} \times {L_{out}}}}$ is log local features. We use a convolutional kernel of size 3 with ${d_c}$ channels, and ${L_{out}}$ represents the length of the sequence after convolution.

Furthermore, anomalies in microservice systems are not always manifested as short-term events. Some anomalies arise from long-term accumulation processes, such as sustained resource consumption or gradually increasing workload. In such cases, related log events may be temporally distant but still exhibit semantic correlations. To capture these long-range dependencies, CMT-AD incorporates a Transformer encoder to model global contextual relationships:

\begin{equation}
Z_t^g = {\mathop{\rm TransEncoder}\nolimits}(h_t^g)
\label{eq9}
\end{equation}
where $Z_t^g \in {{\mathbb R}^{{L_{out}} \times {d_c}}}$ denotes the log representation produced by the encoder and is used as the input to the decoder, capturing both local features and global contextual information of the log sequence. To facilitate subsequent cross-modal interaction, we set ${d_c} = {d_m}$, so that the hidden dimension of the log representations matches that of the metric representations. Subsequently, the same Transformer encoder structure, followed by an MLP layer, is used for reconstruction, as shown in Eq.~\ref{eq10}. This design avoids potential representation shifts caused by inconsistent encoder-decoder architectures while maintaining reconstruction capability with reduced modeling complexity. 

\begin{equation}
\hat X_t^g = {\mathop{\rm MLP}\nolimits}({\mathop{\rm TransEncoder}\nolimits}(Z_t^g))\label{eq10}
\end{equation}
where $\hat X_t^g \in {{\mathbb R}^{s \times n1}}$ represents a reconstructed log feature. We also use the mean squared error loss function to calculate the log reconstruction error, as follows:
\begin{equation}
er{r^g} = \sum\limits_{t = 1}^n {{{(X_t^g - \hat X_t^g)}^2}} 
\label{eq11}
\end{equation}

\subsection{Confidence-guided Cross-modal Knowledge Transfer}
In this section, we design a confidence-guided cross-modal knowledge transfer module that quantifies modality reliability as confidence scores. Under dynamically varying modality confidence levels, the proposed mechanism facilitates knowledge transfer from high-confidence modalities to low-confidence ones, thereby producing more stable multimodal representations for anomaly detection.

\subsubsection{Modality Confidence Estimation via Soft Clustering Assignment}
Considering the high cost of acquiring labeled data, we use the soft assignment results from deep clustering to assess the reliability of different modalities. Specifically, given the metric and log representations under the $t$-th window as $Z_t^m$ and $Z_t^g$, respectively, we map them into a unified clustering feature space and obtain the soft clustering results $q_{tc}^m$ and $q_{tc}^g$ based on the cluster centers $\{ {\mu _c}\} _{c = 1}^C$. The detailed calculation is introduced in the subsequent anomaly detection process, where $C$ denotes the number of cluster centers, $q_{tc}^m$ and $q_{tc}^g$ represent the probabilities that the metric and log representations under the $t$-th window belong to the $c$-th cluster, respectively. Subsequently, we utilize the entropy of cluster soft assignments as a measure of modality confidence. The intuition is that cluster uncertainty reflects how clearly a modality representation can be assigned to a specific cluster. When a modality provides sufficient information with clear structural patterns, its representation tends to be associated with a specific cluster, resulting in a peaked soft assignment distribution and lower entropy, which indicates higher confidence. Conversely, noisy or less informative modality representations lead to more ambiguous cluster assignments, producing a more uniform soft assignment distribution and higher entropy. Since both metric and log modalities are mapped into the same clustering space, their soft assignment entropy provides an estimation of relative modality reliability, enabling dynamic confidence estimation for cross-modal fusion:
\begin{equation}
c_t^w = 1 - \frac{{H(q_t^w)}}{{\log C}}
\label{eq12}
\end{equation}

\begin{equation}
H(q_t^w) =  - \sum\limits_{c = 1}^C {q_{tc}^w\log q_{tc}^w}
\label{eq13}
\end{equation}
where $w \in \{ m,g\}$ represents the metric or log modal. $c_t^w$ represents the confidence score of the $t$-th sample in mode $w$. $H(q_t^w)$ is the information entropy of the sample in mode $w$. This confidence weighting approximates risk-aware alignment under uncertain representations, suppressing high-entropy samples that may induce negative transfer.

\subsubsection{Confidence-guided Gated Intermediate Modality Construction}
Although confidence scores can indicate the reliability of different modalities at various times, an inherent modal gap exists between metrics and logs. In particular, after introducing confidence scores, directly performing cross-modal knowledge transfer not only fails to effectively leverage the key information from high-confidence modalities but may also amplify unreliable knowledge from low-confidence modalities, thereby weakening the transfer effect. To address this issue, we propose an intermediate modality (IM)~\cite{cmkt} that regulates cross-modal fusion based on modality-specific clustering confidence scores. Specifically, features from each modality are first projected into a shared representation space, and a confidence-guided gating mechanism adaptively weights their contributions. This mechanism enables high-confidence modalities to exert greater influence while suppressing noisy or unreliable signals, thus mitigating cross-modal heterogeneity. The calculation is shown in Eq.~\ref{eq14},~\ref{eq15}.

\begin{equation}
r_t^m = \frac{{c_t^m}}{{c_t^m + c_t^g}},r_t^g = \frac{{c_t^g}}{{c_t^g + c_t^m}}
\label{eq14}
\end{equation}
\begin{equation}
Z_t^{mid} = r_t^g \cdot {F_g}(Z_t^g) + r_t^m \cdot {F_m}(Z_t^m)
\label{eq15}
\end{equation}
where ${F_g}( \cdot )$ and ${F_m}( \cdot )$ are a mode-specific linear projection function. $r_t^g$ and $r_t^m$ represents the confidence score after the gating mechanism update. ${d_{mid}}$ represents the IM dimension. Furthermore, we introduce another MLP layer to enhance the expressive power of ${Z_t^{mid}}$.

\subsubsection{Transfer Stability Rregularization}
Prior to cross-modal knowledge transfer, CMT-AD performs transfer stability regularization from both intra-and cross-modal perspectives. By enhancing the compactness of representations within the same modality clusters and suppressing uncontrolled proximity between different modalities and distinct clusters, the model establishes stable and transferable feature representations for subsequent cross-modal transfer.

\textbf{Intra-modal regularization term:} To stabilize modality-specific representations, we introduce a regularization term guided by the soft cluster assignments. For the intra-modal representation $Z_t^w$, $w \in \{ m,g\}$ of the $t$-th window sample and the soft-assignment weights $q_{ic}^w = \{ q_{ic}^m,q_{ic}^g\}$ in the $c$-th clusters, we first calculate the corresponding soft cluster centers $e_c^w$:
\begin{equation}
e_c^w = \frac{{\sum\limits_{t = 1}^B {q_{tc}^wZ_t^w} }}{{\sum\limits_{t = 1}^B {q_{tc}^w + \varepsilon } }}
\label{eq16}
\end{equation}
where $\varepsilon = 10^{-10}$ is a numerical stabilization term used to avoid division by zero. Subsequently, CMT-AD brings intra-modal sample features from the same cluster closer together and separates those from different clusters:
\begin{equation}
L_{pull}^w = \frac{1}{B}\sum\limits_{t = 1}^B {\sum\limits_{c = 1}^C {q_{tc}^w||Z_t^w - e_c^w||_2^2} }
\label{eq17}
\end{equation}

\begin{equation}
L_{push}^w = \frac{1}{{C(C - 1)}}\sum\limits_{c \ne c'} {\exp ( - ||e_c^w - e_{c'}^w||_2^2)}
\label{eq18}
\end{equation}
It can be observed that if $q_{tc}^w$ is larger in $L_{pull}^w$, meaning the sample is very close to the $c$ then after regularization, it will be pulled toward the cluster center $e_c^w$. $L_{push}^w$ makes different clusters distinguishable in the feature space. Consequently, the intra-modal regularization term is ${L_{tra}} = {L_{pull}} + {L_{push}}$.

\textbf{Cross-modal regularization term:} In complex microservice systems, not all cross-modal interactions contribute positively to anomaly detection, especially when features from different modalities correspond to opposing system states. Direct cross-modal transfer may result in negative transfer. Therefore, CMT-AD mitigates this by penalizing misalignment, increasing the feature distance between samples from different clusters across modalities.
\begin{equation}
{L_{ter}} = \frac{1}{{C(C - 1)}}\sum\limits_{c \ne c'} {\max (0,{\delta _{ter}} - ||e_c^m - e_{c'}^g|{|_2})}
\label{eq19}
\end{equation}
where ${\delta _{ter}}$ is the minimum segmentation interval. Penalty is applied only when $||e_c^m - e_{c'}^g|{|_2} < {\delta _{ter}}$, The overall regularization is implemented as a loss composed of two terms: ${L_{reg}} = {L_{tra}} + {\lambda _{reg}}{L_{ter}}$, ${\lambda _{reg}}$ is used to mitigate the optimization-dominant problem caused by differences in the numerical scales of various regularization terms and to adjust the relative influence of the two types of constraints on model training.

\begin{figure*}[!t]
    \centering
    \includegraphics[width=0.96\textwidth]{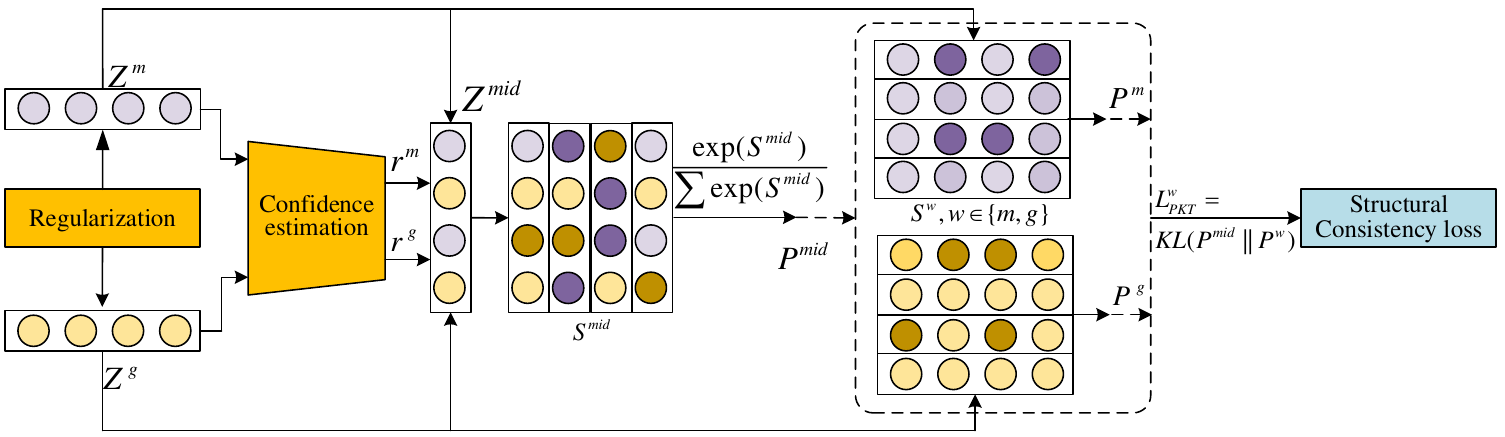}
    \caption{Structural consistency constraint.}
    \label{fig2}
\end{figure*}

\subsubsection{Structural Consistency Constraint}
We recognize that even when different modalities are projected into a unified representation space, effective anomaly patterns may still be weakened during cross-modal knowledge transfer. This is due to inherent differences in how metrics and logs characterize anomalies, including variations in manifestation patterns and discriminative granularity. Therefore, CMT-AD constrains the cross-modal knowledge transfer process by enforcing structural consistency through a dedicated consistency loss. Inspired by existing research on geometric regularization within and between channels~\cite{SemiCMT}, as shown in Fig.~\ref{fig2}, we employ the probabilistic knowledge transfer algorithm PKT~\cite{hong,audio} to align the similarity distributions of samples from different modalities, with the IM serving as a structural reference. Given a mini-batch of size $B$, the IM representations are denoted as${Z^{mid}} = [Z_1^{mid},Z_2^{mid},...,Z_B^{mid}]$, and the unimodality denoted as ${Z^w} = [Z_1^w,Z_2^w,...,Z_B^w]$, $w \in \{ m,g\}$. First, normalization is performed to ensure that the similarities are comparable, resulting in ${\tilde Z^{mid}}$ and ${\tilde Z^{w}}$. Then, the similarity matrix and the relative similarity distribution within the batch are calculated:

\begin{equation}
S_{ij}^{mid} = \tilde Z_i^{mid} \cdot \tilde Z_j^{mid},S_{ij}^w = \tilde Z_i^w \cdot \tilde Z_j^w
\label{eq20}
\end{equation}
where $S_{ij}^{mid}$ represents the similarity matrix of IM. $S_{ij}^{w}$ represents the similarity matrix of a single mode. By calculating the similarity matrix, we model the structure as relationships between samples. The key step in PKT involves computing the relationship between each sample and the other samples in the batch as a probability distribution, denoted as $P_{ij}^{mid}$ and $P_{ij}^{w}$, representing the similarity distributions of the IM and the single modality, respectively. Here, $\gamma $ is an adjustable temperature coefficient. Finally, we use the KL divergence as a constraint to align the similarity distribution of the signal modality samples with that of the IM samples, obtaining the structural consistency loss.
\begin{equation}
L_{PKT}^w = \frac{1}{B}\sum\limits_{i = 1}^B {{\mathop{\rm KL}\nolimits}(P_{i,:}^{mid}||P_{i,:}^w)} 
\label{eq21}
\end{equation}

\begin{equation}
{L_{str}} = (1 - {\tilde r^m})L_{PKT}^m + (1 - {\tilde r^g})L_{PKT}^g\label{eq22}
\end{equation}
where ${\tilde r^w} = \frac{1}{B}\sum\limits_{t = 1}^B {r_t^w}$ is the average gating weight of samples within the batch. In the structure consistency loss, the modality with higher confidence scores serves as a structural reference during cross-modal transfer. By minimizing divergence between similarity distributions, the method limits the geometric structure deviation between samples, thereby reducing representation drift in the shared latent space.

\subsubsection{Semantic Consistency Constraint}
\begin{figure}[htbp]%
		\centering
		\includegraphics[width=2.2in]{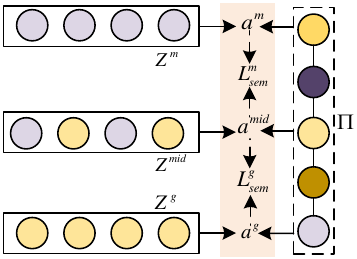}
		\caption{Semantic consistency constraint.}\label{fig3}
	\end{figure}

As shown in Fig.~\ref{fig3}, semantic consistency is enforced by aligning cross-modal distributions within a shared semantic representation space. Specifically, we initialize a set of learnable shared semantic prototypes, denoted as $\Pi = {\pi_1, \pi_2, ..., \pi_M}$, from a standard normal distribution, where $M$ represents the number of prototypes used to model multimodal semantic distributions across metrics, logs, and intermediate modalities. A smaller $M$ may limit the expressiveness of semantic variations, while a larger $M$ may introduce redundancy and increase optimization difficulty. We empirically search $M \in {4, 8, 16, 32, 64}$ and find that $M=16$ achieves a good trade-off between representation capacity and computational efficiency. The prototypes are then normalized to obtain $\tilde{\Pi}$, and cosine similarity is computed between IM and unimodal representations and the prototypes to derive modality-specific semantic distributions.


\begin{equation}
s_{ij}^{w'} = \frac{{\exp (a_{ij}^{w'})}}{{\sum\limits_{\ell  = 1}^M {\exp (a_{i\ell }^{w'})} }}
\label{eq23}
\end{equation}
where $w' = \{ m,g,mid\}$, $a_{ij}^{w'}$ can be calculated from the similarity score between the $i$-th sample and the $j$-th prototype, and $s_{ij}^{w'}$ denotes the corresponding semantic similarity distribution. Therefore, for each sample in the current batch, we obtain a probability vector $s_i^{w'} = [s_{i1}^{w'},s_{i2}^{w'},...,s_{iM}^{w'}]$ of length $M$. Finally, we apply the KL divergence constraint to align the semantic similarity distribution of the metrics and logs with that of the IM, obtaining the semantic consistency loss.
\begin{equation}
L_{sem}^w = \frac{1}{B}\sum\limits_{i = 1}^B {{\mathop{\rm KL}\nolimits}(s_i^{mid}||s_i^w)}\label{eq24}
\end{equation}

\begin{equation}
{L_{sem}} = (1 - {\tilde r^m})L_{sem}^m + (1 - {\tilde r^g})L_{sem}^g
\label{eq25}
\end{equation}
Similar to structural consistency constraint, semantic consistency also uses the IM as a reference and reduces the dispersion of the semantic space by introducing a shared prototype mechanism, thereby stabilizing the cross-modal shared semantic cluster center.

\subsection{Anomaly Detection}
We employ the simple yet effective K-means algorithm to cluster and differentiate various operating modes of the system~\cite{mad-cmc}. Additionally, it optimizes the clustering and feature extraction processes simultaneously to achieve accurate clustering results~\cite{cross}. Specifically, in the confidence-guided cross-modal knowledge transfer module, we obtain the IM representation $Z_t^{mid}$ which integrates metrics and log information, as input to the K-means algorithm. We then determine $C$ cluster centers $\{ \mu \} _{c = 1}^C$ and apply a distance-based soft assignment strategy to calculate the probability that each sample belongs to a particular cluster:

\begin{equation}
{q_{tj}} = \frac{{{{(1 + ||Z_t^{mid} - {\mu _j}|{|^2})}^{ - 1}}}}{{\sum\limits_{c = 1}^C {{{(1 + ||Z_t^{mid} - {\mu _c}|{|^2})}^{ - 1}}} }}\label{eq26}
\end{equation}
where ${q_{tj}}$ represents the probability that the $t$-th window sample is assigned to $j$-th cluster. Subsequently, to enhance clustering accuracy and reduce the influence of low-confidence assignments, we construct an auxiliary target distribution $P = \{ {p_{tj}}\}$ to emphasize high-confidence cluster assignments and optimize the clustering process, defined as follows:
\begin{equation}
{p_{tj}} = \frac{{q_{tj}^2/{f_j}}}{{\sum\limits_{c = 1}^C {q_{tc}^2/{f_c}} }},{f_j} = \sum\nolimits_t {{q_{tj}}}\label{eq27}
\end{equation}
where ${f_j}$ represent the number of soft samples in the current batch for the $j$-th cluster. Finally, by minimizing the KL divergence between the predicted distribution $Q$ and the target distribution $P$, a deep clustering loss ${L_{clu}} = {\mathop{\rm KL}\nolimits}(P||Q)$ is formulated to encourage the model to create compact intra-cluster representations and well-separated inter-cluster representations, thereby improving the distinction between different system states.

The overall model loss includes clustering loss, regularization, structural and semantic consistency constraints, calculated as follows:
\begin{equation}
L = {L_{clu}} + \alpha {L_{reg}} + \beta {L_{str}} + \lambda {L_{sem}}
\label{eq28}
\end{equation}
where $\alpha$, $\beta$ and $\lambda$ are three balancing factors used to adjust different loss weights. It is worth noting that CMT-AD does not perform clustering directly in a randomly initialized representation space. The metric and log encoders are first independently pre-trained for a maximum of 30 epochs using reconstruction objectives. After pretraining, the selected encoder parameters are used to initialize the cross-modal training stage, and K-means is applied to the pretrained representations to initialize the cluster centers $\{ \mu \} _{c = 1}^C$. During joint training, the cluster centers are treated as trainable parameters and updated together with the network parameters through backpropagation by minimizing the clustering loss.

After model convergence, the samples are divided into \(C=3\) clusters, corresponding to the normal, abnormal, and transitional states. To obtain a reproducible state mapping, we characterize each cluster from three aspects: sample proportion \(R_j\), representation dispersion \(D_j\), and soft-assignment uncertainty \(H_j\). For the \(j\)-th cluster, these quantities are defined as follows:
\begin{equation}
\begin{aligned}
{R_j} = \frac{{{n_j}}}{n},{D_j} = \frac{1}{{{n_j}}}\sum\limits_{t \in {{{\cal C}}_j}} {||Z_t^{mid} - {\mu _j}|{|_2}},\\
{H_j} = -\frac{1}{{{n_j}}}\sum\limits_{t = 1}^{{n_j}} {\frac{{\sum\limits_{c = 1}^C {{q_{tc}}\log {q_{tc}}} }}{{\log C}}}
\end{aligned}
\label{eq29}
\end{equation}
where $n_j$ denotes the number of samples in the $j$-th cluster ${{{\cal C}}_j}$, $n$ is the total number of samples, $\mu _j$ denotes the corresponding cluster center, and $q_{tc}$ is the soft-assignment probability of sample $t$ to cluster $c$. A larger \(R_j\) indicates that the cluster contains more samples, while larger \(D_j\) and \(H_j\) indicate greater representation dispersion and assignment uncertainty, respectively.

Since normal states generally contain more samples with compact and stable representations, whereas transitional states tend to exhibit greater dispersion and uncertainty, we define the normal-state score and transitional-state score as:
\begin{equation}
S_j^{nor} = {\tilde R_{j}} + (1 - {\tilde D_j}) + (1 - {H_j}), S_j^{tra} = {\tilde D_j} + {H_j}
\label{eq30}
\end{equation}
where $\tilde R_{j}$ and ${\tilde D_j}$ are obtained by min-max normalization across the \(C\) clusters. The cluster with the highest $S_j^{nor}$ is identified as the normal state. Among the remaining two clusters, the one with the higher \(S_j^{tra}\) is identified as the transitional state, while the remaining cluster is regarded as the abnormal state. For each sample in the transitional cluster, we further calculate the Euclidean distances between its IM feature and the normal and abnormal cluster centers, denoted by \(d_{nor}\) and \(d_{abnor}\), respectively. If \(d_{nor}>d_{abnor}\), the sample is classified as anomalous; otherwise, it is classified as normal. Finally, all samples are assigned their final labels to obtain the anomaly detection results.

\section{Experiments}
\subsection{Experimental Design and Implementation}
\subsubsection{Dataset}
\textbf{The GAIA dataset} is an open-source benchmark developed by Cloudwise for AIOps research.\footnote{https://github.com/CloudWise-OpenSource/GAIA-DataSet} It was collected from a microservice system environment comprising physical hosts and 10 microservices, each monitoring 83 metric variables. In total, the dataset includes over 700,000 metric data points and 87 million log data points.

\textbf{The Spark dataset} released by the Hades study, is designed to analyze the state of microservice systems using metric and log data.\footnote{https://zenodo.org/records/7609780} Spark 3.0 was deployed in a distributed cluster comprising one master node and five worker nodes. A total of 19 workloads were executed to cover diverse service scenarios, monitoring 11 metric variables. The complete dataset contains over 60,000 metric records and approximately 1 million log entries. 

\textbf{The Stack dataset}~\cite{msds} originates from the complex distributed system OpenStack and is used for AI-powered analytics.\footnote{https://zenodo.org/record/3549604} It is collected from five physical nodes within the technical facility, comprising over 400,000 metric data points and more than 160,000 logs, encompassing 25 metric variables. After log parsing, the dataset includes 216 log templates, with a normal to abnormal data ratio of approximately 80:1.

\subsubsection{Baseline}
We compare CMT-AD with several baseline methods, including metric-based methods: Anomaly-Transformer~\cite{anomalytrans} and MUTANT~\cite{mutant}, log-based methods: DeepLog~\cite{deeplog}, LightLog~\cite{lightlog} and DeepSysLog~\cite{deepsyslog}, and multimodal anomaly detection methods: SCWarn~\cite{scwarn}, UAC-AD~\cite{uac-ad}, and KANAD~\cite{kanad}.

\subsubsection{Evaluation Measurements}

We use three widely used evaluation metrics to measure the performance of CMT-AD and other methods: $Precision = \frac{{TP}}{{TP + FP}}$, $Recall = \frac{{TP}}{{TP + FN}}$ and $F1 - score = \frac{{2*Precision*Recall}}{{Precision + Recall}}$. Where $TP$ represents the number of abnormal samples correctly detected, $FP$ represents the number of normal samples incorrectly classified as abnormal, and $FN$ represents the number of abnormal samples incorrectly classified as normal.

\subsubsection{Implementation}
Our experiments are conducted on a server equipped with an NVIDIA GeForce RTX 3090 GPU with 24 GB of GPU memory, using Python 3.8, PyTorch 1.9.0, and CUDA 11.1. The dataset is divided into three parts: a training set (70\%), a validation set (10\%), and a test set (20\%). For the sliding window, we set the window size $T$ to 20 and the step size to 1. Regarding metrics, considering the number and complexity of services in various microservice systems, we set $K=25$ in the GAIA dataset to construct feature graphs, $K=5$ in the Spark dataset, and $K=11$ in the Stack dataset. Regarding logs, we use the off-the-shelf service bert-as-service to obtain log semantics. During SVD dimensionality reduction, we calculate the number of dimensions required to retain over 95\% of the semantic information across different datasets, resulting in approximately 110–130 dimensions. To unify the model input dimension, the dimensionality of the log semantic features is ultimately fixed at 128. We set the hidden layer dimensions ${d_m} = {d_c} = 32$, the regularization balance factor $\alpha$ to 0.06, the structural consistency constraint balance factor $\beta$ to 0.2, the semantic consistency constraint balance factor $\lambda$ to 0.05, the number of clusters $C=3$, the batch size to $B=64$, and Adam is used as the optimizer with an initial learning rate of 0.001. We set the maximum number of epochs to 100 and implemented an early stopping strategy to dynamically terminate training. Regarding the baselines, we determined the hyperparameters based on the publicly available implementations. All performance experiments were independently repeated five times using random seeds \(\{0,1,2,3,4\}\), and the results are reported as mean \(\pm\) standard deviation. The source \footnote{https://github.com/wpp33669-hub/CMT-AD} are publicly
available.

\subsection{The Overall Performance of CMT-AD}
The overall experimental results are presented in Table~\ref{tab2}. It can be observed that CMT-AD achieves the highest F1-score in the three datasets, 0.975, 0.911 and 0.967 respectively. This indicates that the CMT-AD demonstrates superior accuracy in identifying anomaly samples. When comparing the two metric-based methods, MUTANT's F1-score is 12.5\%, 2.4\% and 19\% higher than AnoTrans's in the respective datasets. This improvement is attributed to MUTANT's ability to analyze relationships between different variables through sequence modeling, effectively capturing the temporal-spatial features of the metrics. These findings further validate the advantage of CMT-AD in modeling spatial features.

Among the three log-based methods, the Precision and Recall of DeepSysLog and LightLog are higher than DeepLog. This improvement is mainly because they convert logs into semantic sequences, which better reflect log anomalies and are more robust to noise compared to index-based methods. DeepSysLog further incorporates timestamps, IP addresses, and other metadata, capturing richer semantic embeddings. Therefore, CMT-AD employs BERT to extract semantic log features.

Among multimodal anomaly detection methods, SCWarn performed the worst, even lagging behind some single-modal anomaly detection approaches on three datasets. This underperformance can be attributed to two main factors: first, SCWarn relies solely on a temporal model to analyze both metrics and logs simultaneously, neglecting the spatial features of metrics and the semantic representations of logs; second, it does not account for the interactions between metrics and logs, instead depending exclusively on a parallel architecture for separate predictions. In contrast, UAC-AD achieved the best performance among all comparative methods on the GAIA dataset, with an F1-score of 0.951. Its advantage stems not only from employing an attention mechanism to facilitate cross-modal interaction but also from utilizing contrastive learning and adversarial training to enhance the model's discriminative ability on hard samples within the dataset.

Although KANAD's F1-score is lower than UAC-AD's on the GAIA dataset, its F1-scores on the Spark and Stack datasets are 1.6\% and 1.9\% higher, respectively. All three datasets show F1-scores exceeding 0.9. This approach not only uses KNN to model the metric graph but also constructs a log semantic graph using pointwise mutual information and employs contrastive learning to achieve cross-modal interaction and alignment. However, compared with CMT-AD, both KANAD and UAC-AD implicitly assume fixed contributions from metrics and logs without dynamically adjusting their importance. In contrast, CMT-AD not only learns multi-perspective representations for each modality but also adaptively enhances low-confidence modality representations through confidence-guided cross-modal transfer learning, thereby achieving superior detection performance.

\begin{table*}[!t]
\caption{ANOMALY DETECTION PERFORMANCE}
\label{tab2}
\centering
\renewcommand{\arraystretch}{1.2}
\begin{tabular}{|c|ccc|ccc|ccc|}
\hline
\multirow{2}{*}{Methods} 
& \multicolumn{3}{c|}{GAIA} 
& \multicolumn{3}{c|}{Spark} 
& \multicolumn{3}{c|}{Stack} \\
\cline{2-10}
& Precision & Recall& F1-score& Precision& Recall & F1-score & Precision & Recall& F1-score \\
\hline

AnoTrans & 0.453$\pm$0.002 & 0.688$\pm$0.019 & 0.546$\pm$0.003 & 0.628$\pm$0.002 & 0.839$\pm$0.009 & 0.718$\pm$0.005 & 0.551$\pm$0.045 & 0.682$\pm$0.043 & 0.610$\pm$0.045 \\
MUTANT   & 0.679$\pm$0.001 & 0.664$\pm$0.089 & 0.671$\pm$0.002 & 0.671$\pm$0.014 & 0.830$\pm$0.049 & 0.742$\pm$0.019 & 0.785$\pm$0.112 & 0.816$\pm$0.060 & 0.800$\pm$0.062 \\
DeepLog   & 0.364$\pm$0.018 & 0.681$\pm$0.021 & 0.474$\pm$0.010 & 0.218$\pm$0.038 & 0.409$\pm$0.034 & 0.284$\pm$0.040 & 0.490$\pm$0.019 & 0.611$\pm$0.029 & 0.544$\pm$0.021 \\
LightLog  & 0.778$\pm$0.001 & 0.837$\pm$0.007 & 0.806$\pm$0.003 & 0.340$\pm$0.007 & 0.699$\pm$0.011 & 0.457$\pm$0.004 & 0.857$\pm$0.008& 0.829$\pm$0.015& 0.843$\pm$0.012 \\
DeepSysLog& 0.785$\pm$0.012 & 0.850$\pm$0.025 & 0.816$\pm$0.011 & 0.486$\pm$0.027 & 0.773$\pm$0.035 & 0.597$\pm$0.031 & 0.845$\pm$0.021 & 0.889$\pm$0.019& 0.866$\pm$0.002 \\
SCWarn & 0.482$\pm$0.018 & 0.651$\pm$0.011 & 0.553$\pm$0.012 & 0.289$\pm$0.002 & 0.382$\pm$0.006 & 0.329$\pm$0.005 & 0.446$\pm$0.032 & 0.350$\pm$0.030 & 0.397$\pm$0.022  \\
UAC-AD & 0.955$\pm$0.014 & 0.947$\pm$0.014 & 0.951$\pm$0.014 & \textbf{0.892$\pm$0.002} & 0.890$\pm$0.004 & 0.891$\pm$0.004 & 0.885$\pm$0.008 & \textbf{1.000$\pm$0.000}&0.939$\pm$0.008 \\
KANAD     & 0.920$\pm$0.006 & 0.935$\pm$0.002 & 0.927$\pm$0.001 & 0.886$\pm$0.005 & 0.929$\pm$0.024 & 0.907$\pm$0.005 & 0.936$\pm$0.003 & 0.980$\pm$0.020 & 0.958$\pm$0.003 \\
CMT-AD    & \textbf{0.969$\pm$0.001} & \textbf{0.982$\pm$0.014} & \textbf{0.975$\pm$0.007} &
          0.891$\pm$0.017 & \textbf{0.933$\pm$0.056} & \textbf{0.911$\pm$0.034}
          & \textbf{0.949$\pm$0.032} & 0.985$\pm$0.028 & \textbf{0.967$\pm$0.030} \\
\hline
\end{tabular}
\end{table*}

\subsection{Hyperparameter Sensitivity and Stability Analysis}
\begin{figure*}[t]
\centering

\subfloat[Sliding window size]{
  \includegraphics[trim=26 12 30 35,clip, width=0.31\textwidth]{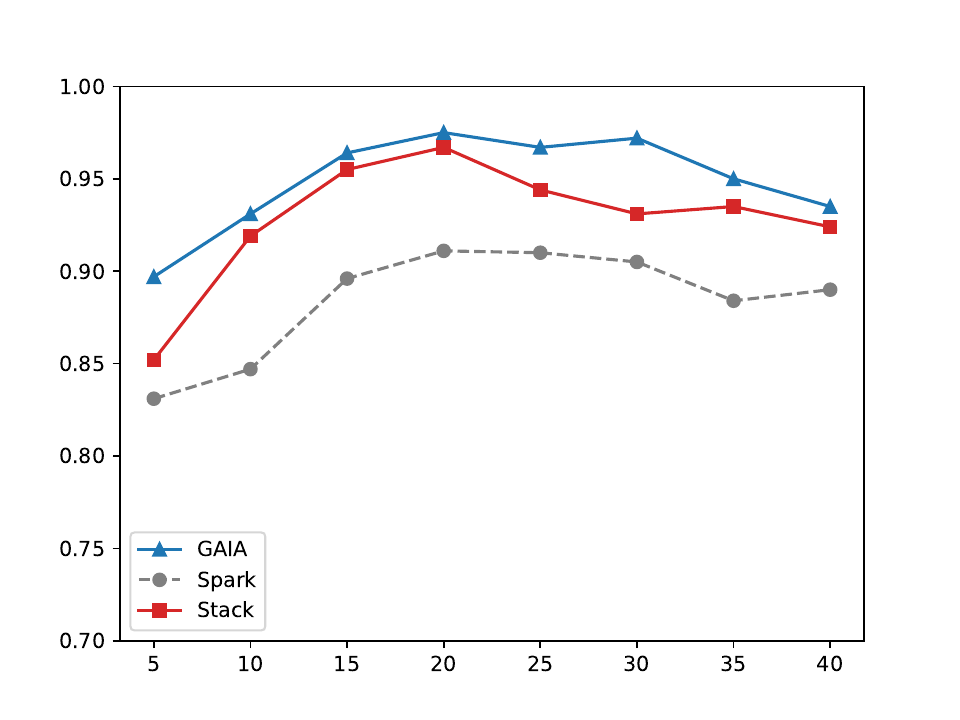}
  \label{fig:4a}
}\hfill
\subfloat[Number of clusters]{
  \includegraphics[trim=26 12 30 35,clip, width=0.31\textwidth]{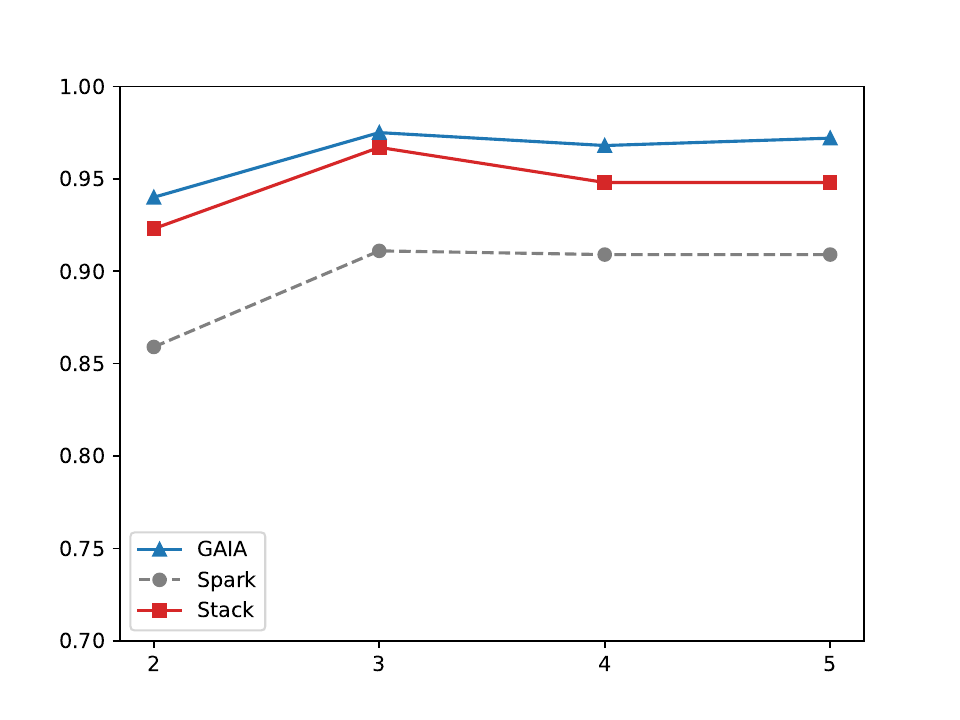}
  \label{fig:4b}
}\hfill
\subfloat[Hidden layer size]{
  \includegraphics[trim=26 12 30 35,clip, width=0.31\textwidth]{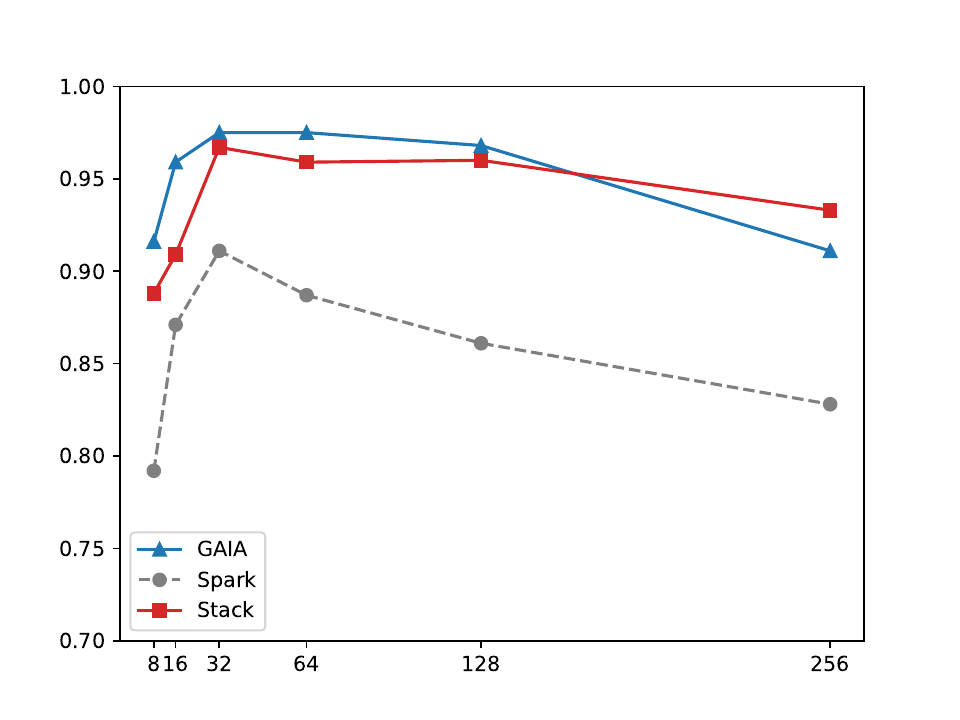}
  \label{fig:4c}
}

\vspace{1.5mm}

\subfloat[Weighting factor of regularization terms $\alpha$]{
  \includegraphics[trim=26 12 30 35,clip, width=0.31\textwidth]{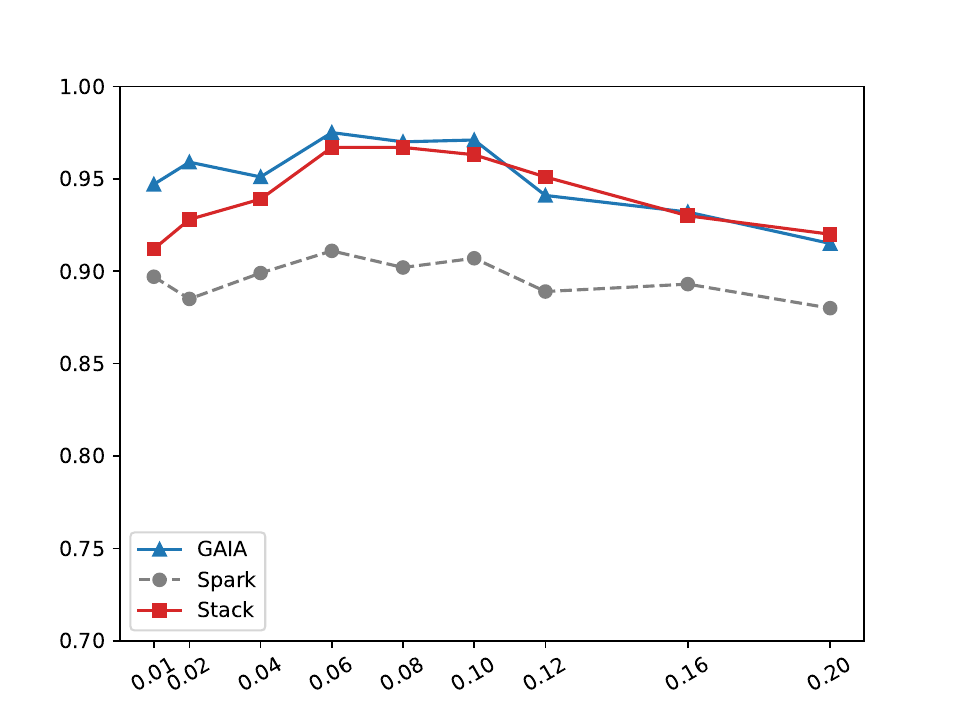}
  \label{fig:4d}
}\hfill
\subfloat[Weighting factor of structural constraint]{
  \includegraphics[trim=26 12 30 35,clip, width=0.31\textwidth]{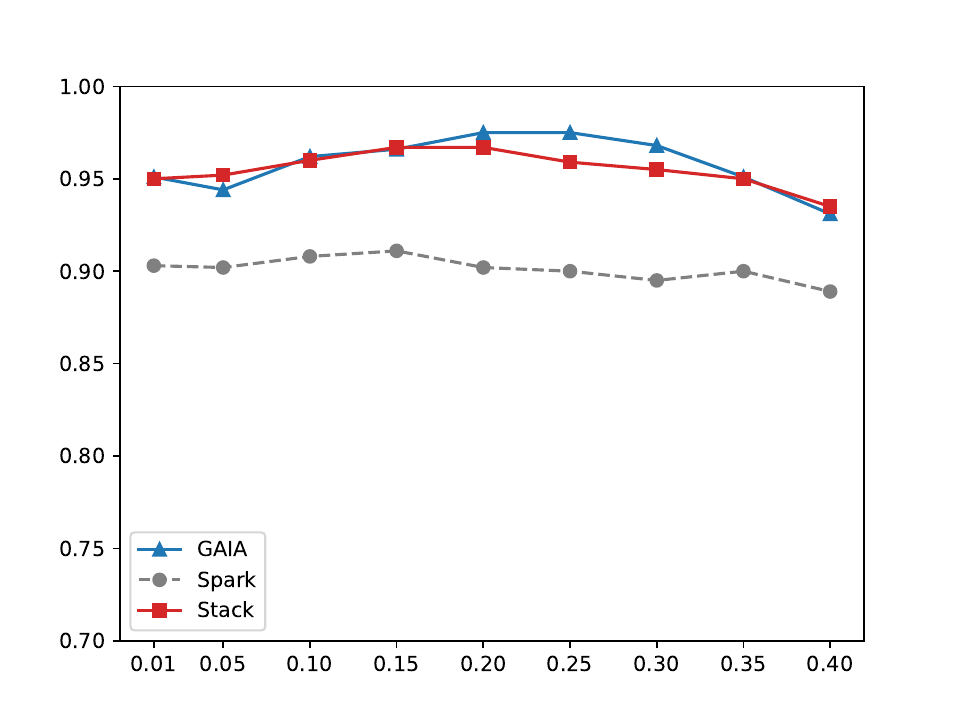}
  \label{fig:4e}
}\hfill
\subfloat[Weighting factor of semantic constraint]{
  \includegraphics[trim=26 12 30 35,clip, width=0.31\textwidth]{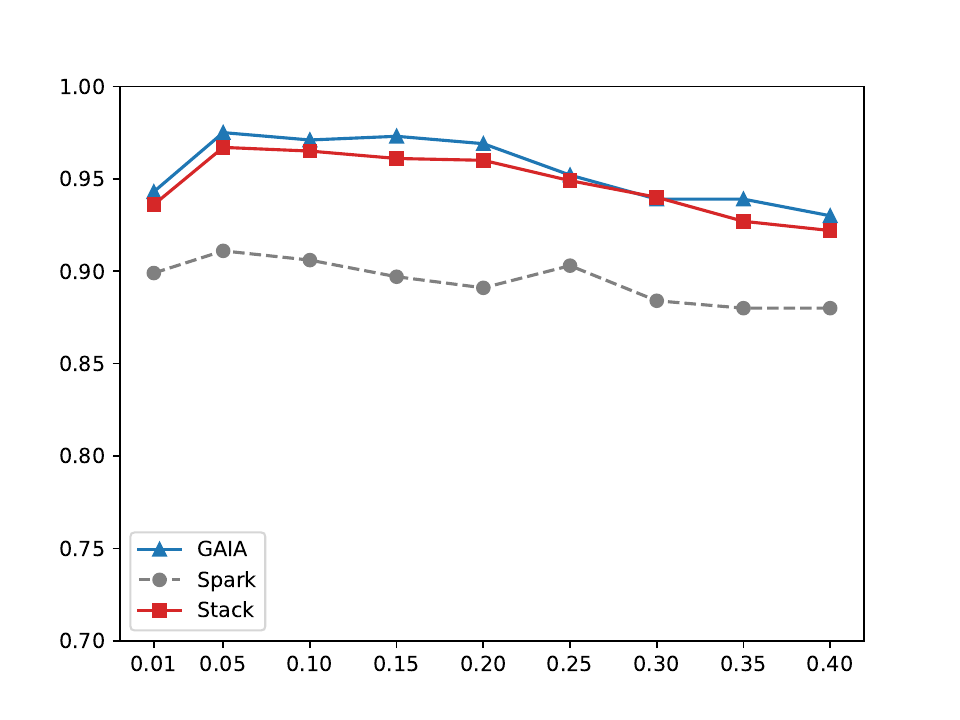}
  \label{fig:4f}
}\hfill
\makebox[0.32\textwidth]{}   

\caption{Anomaly detection results with different parameters.}
\label{fig4}
\end{figure*}

In this section, we analyze the impact of key hyperparameters on CMT-AD performance, including the sliding window size, number of clusters, hidden layer size, and balance factors for regularization and transfer constraints. We also evaluate the stability of the state-mapping strategy under different anomaly ratios. Fig.~\ref{fig4} shows the F1-score results on the three datasets.

As shown in Fig.~\ref{fig:4a}, detection performance is lowest on the three datasets when the sliding window size is smaller than 5. This is because the analysis of the temporal features of the metrics and the global features of the logs is insufficient. As the sliding window size increase, overall detection performance improves, albeit with slight fluctuations. When the window size ranges between 15 and 30, the impact on the GAIA dataset is less pronounced than on the Spark and Stack datasets. This indicates that GAIA is relatively stable over time, resulting in lower sensitivity of the model to window size. Overall, detection performance was optimal when the window size is 20, as this window effectively covered the anomaly evolution cycle while avoiding the introduction of irrelevant information.

As shown in Fig.~\ref{fig:4b}, when $C=2$, the F1-score decreases by at least 3\% across the three datasets. This is because microservice anomalies usually involve transitional states caused by anomaly evolution and slight performance fluctuations, which cannot be effectively distinguished with only normal and abnormal clusters. When $C=3$, these intermediate states can be explicitly modeled, resulting in more stable anomaly decisions. Increasing the number of clusters beyond 3 does not bring significant performance improvements but introduces additional computational overhead. Nevertheless, the optimal number of clusters may vary depending on the complexity of system behaviors and anomaly patterns. In practice, systems with simpler operational behaviors and fewer anomaly patterns may adopt fewer clusters, whereas systems with more complex behaviors and diverse anomaly evolution patterns may require additional clusters to distinguish different intermediate states.

As shown in Fig.~\ref{fig:4c}, the performance initially improves with increasing hidden layer dimensions and reaches the best results at 32. When the hidden layer dimension exceeds 128, a noticeable performance degradation is observed. This decline occurs because larger hidden dimensions increase the model capacity but may also introduce redundant representations and make the model more sensitive to subtle fluctuations or noise. Conversely, an excessively small hidden dimension limits the representation capability, resulting in insufficient modeling of complex multimodal information.

Fig.~\ref{fig:4d} shows that anomaly detection performance deteriorates when the regularization weight is too large. This occurs because excessive regularization over-compresses the feature distribution, forcing the model to focus on consistent representations across modalities while neglecting modality-specific discriminative information. Consequently, the model's ability to distinguish hard samples is weakened. As the regularization weight decreases from a large value, the performance initially improves and reaches its optimum at 0.06. However, further reducing the weight leads to insufficient representation alignment, reducing the stability of transferable modal knowledge and potentially causing negative transfer.

Based on Fig.~\ref{fig:4e} and Fig.~\ref{fig:4f}, the structural and semantic consistency factors are set to 0.2 and 0.05, respectively. A relatively larger structural consistency factor promotes geometric alignment across modalities, while a smaller semantic consistency factor avoids excessive semantic smoothing of anomalous representations. When the consistency factors are too large, the consistency losses may dominate the optimization and weaken the effect of the clustering objective; when they are too small, cross-modal interaction becomes insufficient, both leading to degraded detection performance.

Then, we evaluate the stability of the state-mapping strategy under different anomaly ratios. Since this experiment focuses on isolating the effect of anomaly proportion on state mapping rather than comparing dataset-specific characteristics, GAIA is used as a representative dataset for this stability analysis. Specifically, the trained CMT-AD model is kept fixed, and 10,000 normal samples are retained. We then vary the number of anomalous samples to construct anomaly ratios of 10\%, 20\%, 30\%, and 40\%, respectively, and evaluate the resulting anomaly detection performance.

As shown in Table~\ref{tab6}, CMT-AD maintains stable detection performance as the anomaly ratio increases from 10\% to 40\%. Precision remains nearly unchanged, while Recall improves from 0.957 to 0.980. This indicates that the proposed state-mapping strategy consistently distinguishes normal, abnormal, and transitional clusters across varying anomaly proportions. The slight improvement in Recall at higher anomaly ratios may be attributed to a more comprehensive representation of anomalous patterns, which facilitates easier differentiation of abnormal states within the clustering space. Additionally, these results suggest that the representations learned through cross-modal knowledge transfer remain highly discriminative across different anomaly ratios, providing a reliable feature basis for subsequent state mapping.

\begin{table}[t]
\centering
\caption{STABILITY UNDER DIFFERENT ANOMALY RATIOS ON GAIA}
\label{tab6}
\setlength{\tabcolsep}{8pt}
\renewcommand{\arraystretch}{1.1}
\begin{tabular}{|c|c|c|c|}
\hline
Anomaly Ratio & Precision & Recall & F1-score \\
\hline
10\% & 0.968 & 0.957 & 0.962 \\
20\% & 0.967 & 0.972 & 0.969 \\
30\% & 0.969 & 0.980 & 0.974 \\
40\% & 0.968 & 0.980 & 0.974 \\
\hline
\end{tabular}
\end{table}

\subsection{Ablation Experiment Analysis}
We conduct guided ablation experiments to verify the effectiveness of each module in CMT-AD, including: 1) CMT-AD/m, removing metric data; 2) CMT-AD/g, removing log data; 3) CMT-AD/reg, removing the regularization module; 4) CMT-AD/Stru, removing the structural consistency constraint; 5) CMT-AD/Sem, removing the semantic consistency constraint; 6) CMT-AD/IM, replacing the confidence-guided gating mechanism with average fusion to construct the intermediate modality; 7) CMT-AD/Np, removing unimodal pretraining and directly performing joint training; and 8) CMT-AD/Dr, introducing a 10-epoch warm-up stage after unimodal pretraining, during which only the clustering loss is optimized before activating the regularization and consistency constraints.

Additionally, to further validate the relationship between cross-modal knowledge transfer and unimodal representation quality, we introduce an enhanced variant, CMT-AD++. For metrics, based on an LSTM encoder for temporal feature extraction, we introduce the Fast Fourier Transform (FFT) and concatenate it with temporal features to enrich metric representations from the frequency domain perspective. For logs, we construct an execution relation graph within a sliding window, using normalized frequencies as node embeddings, with adjacency matrix reconstruction as an additional pre-training objective. We then concatenate the graph features with the semantic features generated by the Transformer encoder to enhance log representations. 

The experimental results are shown in Table~\ref{tab3}. When modal ablation is conducted on CMT-AD, i.e., when either metrics or logs are removed, the cross-modal knowledge transfer module is no longer applicable, reducing the regularization to intra-modal regularization. This results in a significant decrease in anomaly detection performance. This finding demonstrates that CMT-AD can identify more anomalies through cross-modal analysis and interaction between metrics and logs. Furthermore, compared to other variant models, CMT-AD/m and CMT-AD/g exhibit lower F1-scores, underscoring the importance of multimodal collaborative analysis for anomaly detection tasks.

Further comparison with advanced single modal methods reveals that CMT-AD/g outperforms MUTANT. This is mainly attributed to the incorporation of intra-modal regularization on top of spatiotemporal feature modeling, which effectively enhances the stability and robustness of metric representations. Compared to DeepSysLog, CMT-AD/m achieves a slightly lower F1-score, mainly because DeepSysLog additionally models metadata information within log variables, thereby obtaining richer semantic features. However, CMT-AD/m still significantly outperforms LightLog, demonstrating the effectiveness of jointly modeling local and global features. 

When the regularization module is removed, the F1-score of CMT-AD/reg decreases by 2.7\%, 0.5\%, and 1.2\% on the three datasets, respectively. This decline occurs because the regularization mechanism suppresses overfitting and improves the stability of single-modality representations, ensuring the stability of transferable knowledge across modalities, and limits cross-modal representation shifts by penalizing misalignment. Removing this module can cause distribution shifts in the encoders of each modality, leading to unstable fusion and redundant information in the transferable knowledge. Consequently, the model struggles to effectively leverage complementary information across modalities.

Removing either the structural or semantic consistency constraint decreases the performance of CMT-AD/Stru and CMT-AD/Sem. Specifically, removing structural consistency disrupts the preservation of relational structures across modalities, making the relative distance relationships in the shared representation space less stable and potentially causing some anomalies to be overlooked. Removing semantic consistency eliminates the guidance of shared semantic prototypes, weakening cross-modal semantic alignment and reducing the discriminative separation between normal and anomalous samples. Moreover, CMT-AD/Sem achieves F1-scores that are 1.5\%, 0.4\%, and 1.5\% higher than those of CMT-AD/Stru on the three datasets, respectively, indicating that structural consistency has a relatively greater impact on detection performance.. 

When confidence-guided gating weighting was replaced with average fusion, the F1-score of CMT-AD/IM decreased across three datasets. This decline occurred because the intermediate modalities in CMT-AD can amplify more informative modalities while suppressing unreliable ones. In contrast, CMT-AD/IM assigns equal weights to different modalities, which may dilute informative representations and introduce redundant or noisy information into the shared representation space. As a result, it weakens cross-modal consistency and reuces the discriminative power between normal and anomalous samples.

As shown in Table~\ref{tab3}, the F1-scores of CMT-AD/Np decrease by 0.6\%, 1.0\%, and 0.5\% on the three datasets, respectively, whereas CMT-AD/Dr shows no obvious performance degradation. This indicates that, although delayed regularization can alleviate the impact of unstable cluster assignments during the early stage of joint training, unimodal pretraining provides a more stable representation initialization for subsequent clustering and thereby helps maintain model performance. In contrast, after removing the pretraining stage, randomly initialized feature representations are more susceptible to early clustering fluctuations, which further reduces the stability of subsequent cross-modal knowledge transfer.

Table~\ref{tab3} shows that enhancing unimodal encoders does not consistently improve performance across all datasets. Although CMT-AD++ has a 0.2\% higher F1-score than CMT-AD on GAIA, this may be attributed to the relatively stable structure of the dataset, where enhanced frequency-domain and structural features provide additional discriminative information. However, the performance gain is not monotonic; on some microservice systems (e.g., Spark and Stack), slight degradation is observed. This phenomenon indicates that, in multimodal settings, simply increasing unimodal representation capacity does not directly translate into better anomaly detection performance. More appropriate fusion strategies may be required to preserve the stability of metric temporal–frequency–spatial features, log semantics, and graph structures. CMT-AD instead focuses on adaptively learning the contributions of different modalities to facilitate cross-modal interaction, rather than relying on stronger unimodal representations, demonstrating that its effectiveness mainly stems from the proposed cross-modal knowledge transfer mechanism.

\begin{table*}[!t]
\caption{RESULTS OF ABLATION EXPERIMENTS}
\label{tab3}
\centering
\renewcommand{\arraystretch}{1.2}
\begin{tabular}{|c|ccc|ccc|ccc|}
\hline
\multirow{2}{*}{Methods}
& \multicolumn{3}{c|}{GAIA}
& \multicolumn{3}{c|}{Spark}
& \multicolumn{3}{c|}{Stack} \\
\cline{2-10}
&Precision& Recall &F1-score&Precision &Recall&F1-score&Precision&Recall&F1-score\\
\hline

CMT-AD/m    & 0.776$\pm$0.006 & 0.861$\pm$0.021 & 0.816$\pm$0.006 & 0.525$\pm$0.020 & 0.608$\pm$0.039 & 0.563$\pm$0.025 & 0.836$\pm$0.019 & 0.891$\pm$0.031 & 0.863$\pm$0.025 \\
CMT-AD/g    & 0.670$\pm$0.003 & 0.799$\pm$0.016 & 0.747$\pm$0.008 & 0.700$\pm$0.019 & 0.899$\pm$0.055 & 0.787$\pm$0.036 & 0.828$\pm$0.022 & 0.845$\pm$0.037 & 0.836$\pm$0.016 \\
CMT-AD/IM   & 0.940$\pm$0.003 & 0.972$\pm$0.009 & 0.956$\pm$0.006 & 0.873$\pm$0.013 & 0.926$\pm$0.026 & 0.899$\pm$0.018 & 0.937$\pm$0.022 & 0.978$\pm$0.016 & 0.957$\pm$0.018 \\
CMT-AD/reg  & 0.929$\pm$0.004 & 0.968$\pm$0.021 & 0.948$\pm$0.011 & 0.892$\pm$0.016 & 0.921$\pm$0.060 & 0.906$\pm$0.035 & 0.941$\pm$0.018 & 0.970$\pm$0.029 & 0.955$\pm$0.011 \\
CMT-AD/Stru & 0.931$\pm$0.009 & 0.933$\pm$0.026 & 0.932$\pm$0.017 & 0.861$\pm$0.027 & \textbf{0.941$\pm$0.053} & 0.899$\pm$0.033 & 0.942$\pm$0.033 & 0.938$\pm$0.039 & 0.940$\pm$0.036 \\
CMT-AD/Sem  & 0.936$\pm$0.005 & 0.959$\pm$0.020 & 0.947$\pm$0.007 & 0.880$\pm$0.021 & 0.927$\pm$0.053 & 0.903$\pm$0.036 & 0.937$\pm$0.035 & 0.974$\pm$0.036 & 0.955$\pm$0.028 \\
CMT-AD/Np & 0.958$\pm$0.018 & 0.980$\pm$0.034 & 0.969$\pm$0.026 & 0.881$\pm$0.039 & 0.922$\pm$0.068 & 0.901$\pm$0.052 & 0.945$\pm$0.029 & 0.980$\pm$0.056& 0.962$\pm$0.042 \\
CMT-AD/Dr & 0.966$\pm$0.001 & 0.982$\pm$0.008 & 0.974$\pm$0.004 & 0.884$\pm$0.020 & 0.933$\pm$0.049 & 0.908$\pm$0.034 & 0.947$\pm$0.026& 0.982$\pm$0.022 & 0.964$\pm$0.015 \\

CMT-AD++    & 0.967$\pm$0.004	& \textbf{0.988$\pm$0.010}	& \textbf{0.977$\pm$0.003}	& \textbf{0.894$\pm$0.020}	& 0.927$\pm$0.059	& 0.910$\pm$0.036	& 0.939$\pm$0.032	& 0.984$\pm$0.033	& 0.961$\pm$0.024 \\
CMT-AD      & \textbf{0.969$\pm$0.001} & 0.982$\pm$0.014 & 0.975$\pm$0.007
             & 0.891$\pm$0.017 & 0.933$\pm$0.056 & \textbf{0.911$\pm$0.034}
             & \textbf{0.949$\pm$0.032} & \textbf{0.985$\pm$0.028} & \textbf{0.967$\pm$0.030} \\
\hline
\end{tabular}
\end{table*}

\subsection{Complexity and Efficiency Analysis}
(1) Computational Complexity Analysis. During the pretraining stage, the metric encoder employs LSTM and GCN to extract temporal and spatial features, respectively. For input data containing $k$ metric variables with a window size of $T$, the time complexity of LSTM is $O(kT{d_m}^2)$ and $d_m$ is the hidden layer dimension, while the graph propagation complexity of GCN is $O(|{E_m}|{d_m})$, $E_m$ is the number of edge. For the log encoder, 1D-CNN is adopted to capture local semantic features, with a complexity of $O(n1{k_g}{d_c}^2)$, where $n1$ is the sequence length, $k_g$ is the size of the convolution kernel, $d_c$ is the hidden layer dimension. Transformer is utilized to capture global dependencies, where the main computational cost comes from the self-attention mechanism with a complexity of $O(n{1^2}{d_c})$.

During the cross-modal knowledge transfer stage, confidence estimation has a complexity of $O(BC{d_c})$, $B$ is the batch size. The complexity of IM construction is $O(Bd_c)$. The intra-modal and cross-modal regularization analyze the distance between samples within a batch and the soft cluster center, with a complexity of $O(BC{d_c} + {C^2}{d_c})$. The structural and semantic consistency constraints have complexities of $O(B^2d_c)$ and $O(BMd_c)$, $M$ is the number of semantic prototypes. Therefore, the overall complexity of the cross-modal knowledge transfer stage is: $O(BC{d_c} + B{d_c} + {C^2}{d_c} + {B^2}{d_c} + BM{d_c})$.

(2) Computational Cost Evaluation. As shown in Table~\ref{tab4}, we evaluated the parameter size, training time, and inference (testing) time of various methods. It can be observed that CMT-AD has a moderate parameter size of 2.7 MB. Although methods such as SCWarn and LightLog have fewer parameters and lower training costs, they rely on simpler feature representations and cannot fully exploit complementary information across modalities. Owing to the additional multimodal representation learning and consistency constraints, CMT-AD incurs higher training costs than single-modal anomaly detection methods, while achieving improved detection performance. Compared with the strong baselines UAC-AD and KANAD, CMT-AD achieves a better balance among model size, detection performance, and inference efficiency. Although both baselines have lower training costs, UAC-AD has a slightly larger model size than CMT-AD, while KANAD introduces a larger model size and higher inference overhead. These results demonstrate that CMT-AD maintains acceptable computational efficiency while incorporating cross-modal knowledge transfer, making it suitable for practical microservice anomaly detection scenarios.

(3) Memory Consumption Analysis. As shown in Table~\ref{tab5}, we report the peak CPU and GPU memory usage across three stages: metric learning, log learning, and anomaly detection using both encoders. It can be observed that the two unimodal learning stages exhibit similar resource consumption, while the overall memory overhead increases only slightly after introducing the cross-modal knowledge transfer module. This result demonstrates that CMT-AD achieves effective cross-modal feature fusion with limited additional memory overhead, improving its practicality for deployment in resource-constrained environments.

\begin{table}[t]
\centering
\caption{COMPUTATIONAL EFFICIENCY COMPARISON OF DIFFERENT METHODS}
\label{tab4}
\setlength{\tabcolsep}{5pt}
\renewcommand{\arraystretch}{1.1}
\begin{tabular}{|l |c| c |c |c|}
\hline
\multirow{2}{*}{Methods} & \multirow{2}{*}{Params (MB)} & \multicolumn{3}{c|}{Training/inference time(s)} \\
\cline{3-5}
 & & GAIA & Spark & Stack \\
\hline
AnoTrans   & 0.44  & 77.98/14.28 & 85.04/14.46 & 61.04/12.36 \\
MUTANT     & 0.48  & 29.66/13.93 & 41.75/18.39 & 26.03/13.41 \\
DeepLog    & 0.10  & 19.42/82.81 &  6.99/17.81 & 25.51/23.80 \\
LightLog   & 0.13 & 12.30/0.63  & 12.77/0.42  & 10.52/0.70  \\
DeepSysLog & 1.41  & 31.37/6.85  & 40.18/5.56  & 32.94/6.71  \\
SCWarn     & 0.05  & 1.94/12.27  & 3.89/17.72  & 2.17/13.82  \\
UAC-AD     & 2.85  & 18.53/0.93  & 59.29/2.07  & 19.25/0.89  \\
KANAD      & 4.75  & 13.57/1.82  & 42.31/9.66  & 9.26/2.05   \\
CMT-AD     & 2.70   & 20.37/0.75  & 72.22/1.65  & 26.11/1.08  \\
\hline
\end{tabular}
\end{table}

\begin{table}[t]
\centering
\caption{RESOURCE CONSUMPTION OF DIFFERENT STAGES}
\label{tab5}
\setlength{\tabcolsep}{5pt}
\renewcommand{\arraystretch}{1.1}
\begin{tabular}{|l|c|c|c|c|}
\hline
\multirow{2}{*}{Stage}
& \multirow{2}{*}{Resource}
& \multicolumn{3}{|c|}{Datasets} \\
\cline{3-5}
& & GAIA & Spark & Stack \\
\hline

\multirow{2}{*}{Metric Learning}
& CPU Memory
& 6.21 GiB & 6.37 GiB & 7.57 GiB \\
\cline{2-5}
& GPU Memory
& 2071 MiB & 2069 MiB & 2218 MiB \\
\hline

\multirow{2}{*}{Log Learning}
& CPU Memory
& 6.23 GiB & 6.38 GiB & 7.57 GiB \\
\cline{2-5}
& GPU Memory
& 2120 MiB & 2115 MiB & 2266 MiB \\
\hline

\multirow{2}{*}{CMT-AD}
& CPU Memory
& 6.34 GiB & 6.43 GiB & 7.65 GiB \\
\cline{2-5}
& GPU Memory
& 2161 MiB & 2153 MiB & 2297 MiB \\
\hline
\end{tabular}
\end{table}

\subsection{Case Study}
\begin{figure}[htbp]
\centering
  \subfloat[Multimodal anomaly evolution in a login failure case.]{
    \begin{minipage}[t]{0.5\textwidth}
      \centering
      \includegraphics[width=0.95\textwidth]{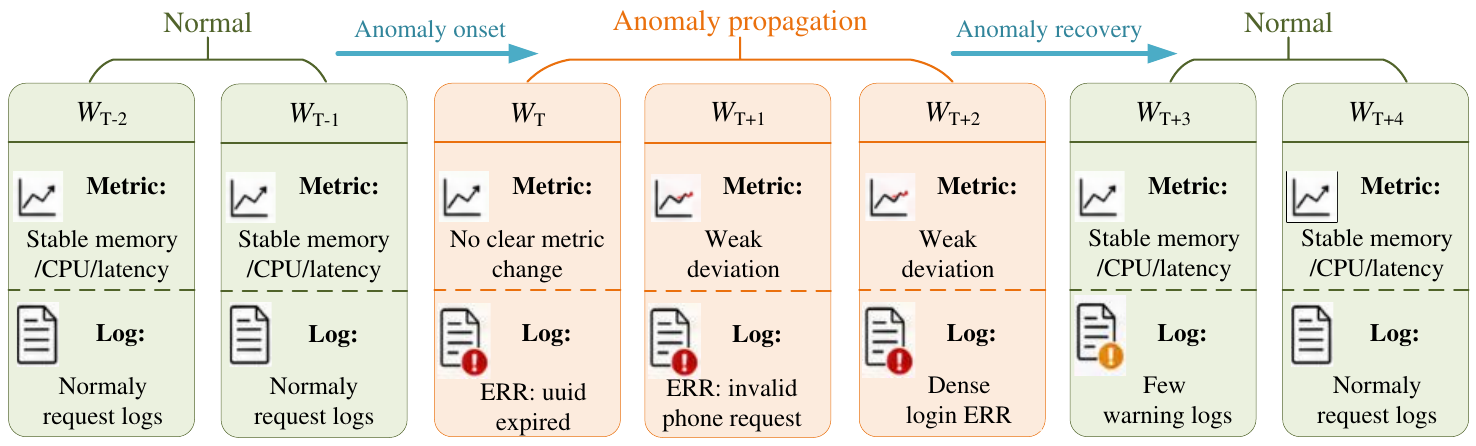}
      \label{5a}
    \end{minipage}
}
  \hfill
  \subfloat[Dynamic confidence changes of metric and log modalities.]{
    \begin{minipage}[t]{0.45\textwidth}
      \centering
      \includegraphics[trim=26 0 0 0,clip,width=0.95\textwidth]{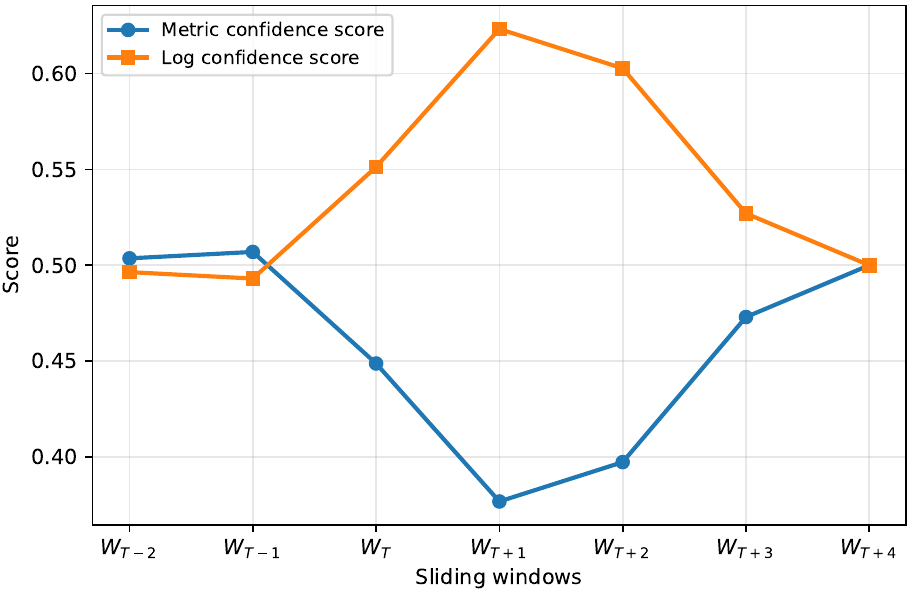}
      \label{5b}
    \end{minipage}
}

  \caption{Case study illustrating the confidence-guided gating mechanism during anomaly evolution.}
  \label{fig5}
\end{figure}

To visually demonstrate the dynamic changes in modality confidence during anomaly evolution, we select a representative "login failure" case from the GAIA dataset. This case is characterized by continuous authentication-related error events in the logs, while the system metrics do not exhibit significant deviations during the anomaly period. As shown in Fig.~\ref{5a}, seven consecutive sliding windows before and after the anomaly are selected to illustrate the corresponding metric and log observations. Fig.~\ref{5b} further presents the dynamic variation of modality confidence scores updated by the confidence-guided gating mechanism.

As shown in Fig.~\ref{5a}, before the anomaly occurrence, both system metrics and logs remain stable, and the confidence scores of the two modalities are relatively balanced. This indicates that both modalities provide consistent and reliable representations of the normal system state. Starting from window $W_T$, authentication failure-related errors, such as "uuid expired" and "invalid phone request", gradually appear in the log stream. However, metrics such as CPU, memory, and latency do not exhibit significant abnormal fluctuations. Therefore, the log modality provides stronger anomaly evidence during this period. As illustrated in Fig.~\ref{5b}, the log confidence score ($c_g$) increases from approximately 0.49 at $W_{T-1}$ to 0.63 at $W_{T+1}$, while the metric confidence score ($c_m$) decreases from approximately 0.51 to 0.38. Based on the changes in reliability of different modalities, the confidence-guided gating mechanism assigns a larger contribution weight to the log modality ($r_g>r_m$), enabling reliable log information to compensate for the insufficient anomaly evidence from metrics. As the system gradually recovers, the number of error-level logs decreases, and the confidence scores of the metric and log modalities return to a balanced state. 

This case demonstrates that, unlike conventional multimodal fusion strategies with fixed contributions, CMT-AD can dynamically adjust modality contributions according to their reliability during anomaly evolution, thereby enhancing the accuracy and robustness of multimodal anomaly detection.

\section{Discussions}
\subsection{Lessons Learned}
(1) Incorporating more modalities does not necessarily lead to improved anomaly detection. Although additional modalities may provide useful information, simple fusion is insufficient to effectively leverage such information, since modality reliability varies during anomaly evolution. Moreover, enhancing single-modal representations does not always synergize with cross-modal interactions. The comparison between CMT-AD++ and CMT-AD shows that additional frequency-domain or graph-structured features may provide supplementary information, but their benefits do not consistently translate into detection gains unless they are effectively integrated into the cross-modal interaction process. Future research should move beyond merely adding more modalities or enhancing single-modal representations, focusing instead on selecting and transferring effective knowledge based on modality reliability.

(2) Cross-modal knowledge transfer is not an unconditional process. Existing studies typically focus on designing more effective alignment and interaction models, such as attention mechanisms and Transformer-based architectures, assuming that information from different modalities is equally suitable for interaction. However, in microservice systems, modality reliability may vary during anomaly evolution, and unconditional transfer may introduce uncertain information. Confidence measures provide a dynamic basis for cross-modal transfer,enabling the model to adaptively select more reliable modality information and avoid performance degradation caused by inappropriate knowledge transfer. Additionally, knowledge transfer based on clustering constraints may be affected by unstable cluster assignments during the early training stage. Single-modal pre-trained encoders and soft clustering strategies provide relatively stable initialization. Nevertheless, under conditions with greater clustering instability, delayed regularization or warm-up strategies become increasingly important for maintaining stable cross-modal knowledge transfer.

\subsection{Threats to Validity}
(1) One potential threat arises from the datasets and data distributions used. We evaluate CMT-AD on three publicly available microservice datasets, which may not fully cover the diverse scenarios in real-world production environments. However, GAIA, Spark, and Stack are collected from microservice systems with different scales and complexities, providing certain representativeness. Furthermore, although the sensitivity analysis selects $C=3$, this setting is motivated by the three typical states in microservice anomaly evolution: normal, transitional, and abnormal. Nevertheless, the clustering-based state identification strategy may still be affected by different anomaly distributions and system complexities. Therefore, the selection of the cluster number could benefit from a more general methodological framework. Future work can incorporate adaptive cluster selection strategies, such as the silhouette coefficient or elbow method, to determine an appropriate number of clusters.

(2) Another potential threat arises from the dynamic evolution of microservice systems. Systems undergo version iterations, service expansions, and configuration changes, resulting in shifts in operating modes, log events, and metric distributions. Such changes may reduce the effectiveness of the learned representations, thereby degrading its performance. However, CMT-AD dynamically adjusts modality contributions according to confidence scores and helps maintain cross-modal representation stability through consistency constraints. Future work could incorporate drift detection, online updates, and incremental learning mechanisms to dynamically adapt model parameters, further enhancing CMT-AD's adaptability in continuously evolving microservice environments.

\section{Conclusions and Future Work}
This paper primarily investigates the effective utilization of reliable cross-modal information from metrics and logs to achieve anomaly detection in microservice systems. To address the reliability challenges posed by dynamic changes in different modalities over time, we design a modal confidence estimation method that enables the model to supplement low-confidence modalities with high-confidence modal knowledge during cross-modal interactions. Simultaneously, to mitigate the impact of cross-modal heterogeneity on knowledge transfer, we conduct an in-depth analysis of the differences in data structures and semantic representations between metrics and logs, and design structural and semantic consistency constraint. Furthermore, we implement intra-modal and cross-modal regularization terms based on clustering soft assignment, ensuring the stability of transferable knowledge and preventing negative transfer by bringing intra-cluster features closer together while pushing inter-cluster features further apart. Extensive experiments on three public datasets consistently demonstrate the superiority of our proposed method.

Our future work will be unfolded along two directions. First, we intend to incorporate anomaly classification and root cause localization following anomaly detection to create a more comprehensive anomaly analysis pipeline. Second, we will investigate shared representations and transfer mechanisms among these three tasks and develop a unified collaborative modeling framework.

\section*{Acknowledgments}
This work is supported in part by the National Natural Science Foundation of China under Grant No. 52231014, and by the Engineering and Physical Sciences Research Council (EPSRC) of UK Research and Innovation (UKRI) under Grant EP/Y028813/1.

 
%
\bibliographystyle{IEEEtran} 
\bibliography{cas-refs}

\end{document}